\documentclass[12pt,notitlepage,a4paper]{article}
\usepackage[utf8]{inputenc}
\usepackage{color,graphicx}
\usepackage{amssymb}
\usepackage{delarray,amsmath,bbm}
\usepackage[american]{babel}
\usepackage{cite}

\def\ie{{\emph{i.e.}}}

\newcommand{\be}{\begin{equation}}
\newcommand{\ee}{\end{equation}}
\newcommand{\bea}{\begin{eqnarray}}
\newcommand{\eea}{\end{eqnarray}}
\numberwithin{equation}{section}

\newfont{\namefont}{cmr10}
\newfont{\addfont}{cmti7 scaled 1440}
\newfont{\boldmathfont}{cmbx10}
\newfont{\headfontb}{cmbx10 scaled 1728}

\begin{document}
\begin{flushright}
HIP-2017-36/TH
\end{flushright}

\baselineskip=15.5pt
\pagestyle{plain}
\setcounter{page}{1}

\begin{center}

\renewcommand{\thefootnote}{\fnsymbol{footnote}}

\begin{center}
\Large \bf Non-commutative massive unquenched ABJM
\end{center}
\vskip 0.3truein
\begin{center}
\bf{Yago Bea,${}^1$\footnote{yagobea@icc.ub.edu}
Niko Jokela,${}^{2,3}$\footnote{niko.jokela@helsinki.fi}
Arttu P\"onni,${}^{2,3}$\footnote{arttu.ponni@helsinki.fi}
and Alfonso V. Ramallo${}^4$\footnote{alfonso@fpaxp1.usc.es}} \\
\end{center}
\vspace{0.5mm}


\begin{center}\it{
${}^1$Departament de F\'\i sica Qu\`antica i Astrof\'\i sica \& Institut de Ci\`encies del Cosmos (ICC), Universitat de Barcelona, Mart\'{\i} i Franqu\`es 1, 08028 Barcelona, Spain
}
\end{center}

\begin{center}\it{
${}^2$Department of Physics and ${}^3$Helsinki Institute of Physics\\
P.O.Box 64 \\
FIN-00014 University of Helsinki, Finland}
\end{center}

\begin{center}\it{
${}^4$Departamento de  F\'\i sica de Part\'\i  culas \\
Universidade de Santiago de Compostela \\
and \\
Instituto Galego de F\'\i sica de Altas Enerx\'\i as (IGFAE)\\
E-15782 Santiago de Compostela, Spain}
\end{center}

\setcounter{footnote}{0}
\renewcommand{\thefootnote}{\arabic{footnote}}

\vspace{0.4in}

\begin{abstract}
\noindent
In this paper we study non-commutative massive unquenched Chern-Simons matter theory using its gravity dual. We construct this novel background by applying a TsT-transformation on the known parent commutative solution. We discuss several aspects of this solution to the Type IIA supergravity equations of motion and, amongst others, check that it preserves ${\cal N}=1$ supersymmetry. We then turn our attention to applications and investigate how dynamical flavor degrees of freedom affect numerous observables of interest. Our framework can be regarded as a key step towards the construction of holographic quantum Hall states on a non-commutative plane.

\end{abstract}

\smallskip
\end{center}

\newpage
\tableofcontents
\newpage

\section{Introduction}

Non-commutative (NC) quantum field theories have a long story in physics \cite{Snider}.  They are non-local and are endowed with a minimal length scale, related to the non-vanishing commutator of the spatial coordinates. The non-commutativity is the origin of the remarkable phenomenon called the UV/IR mixing, in which the ultraviolet and infrared degrees of freedom are mixed in a way similar to what is expected to happen in a theory of quantum gravity. In string theory the NC gauge theories can be realized as D-branes with a Neveu-Schwarz (NS) two-form field $B_2$ \cite{Seiberg:1999vs}.  For these reasons these theories have been studied  extensively in the high-energy physics literature   in the past (for reviews, see \cite{Douglas:2001ba,Szabo:2001kg}).

In the condensed matter physics context, non-commutative geometry appears quite naturally in the study of the quantum Hall effect \cite{Ezawa}. Indeed, after projecting to the lowest Landau level, the position operators of planar electrons in a magnetic field do not commute \cite{GirvinJach} (see also \cite{Ezawa, Comtet:1999owa}). On the other hand, the hydrodynamic models of quantum Hall fluids are described by  a continuous theory containing Abelian Chern-Simons terms, which dominate the long distance dynamics (see, {\emph{e.g.}}, \cite{Fradkin:1991nr,Zee:1996fe,Tong:2016kpv}). Moreover, it was proposed in \cite{Susskind:2001fb} that, in order to take into account the fuzzy ``granular" nature of the electrons, one should consider non-commutative Chern-Simons theory. This non-commutative structure  shows up when the coordinates of the electrons are promoted to the eigenvalues of $N\times N$ Hermitean matrices. Different extensions of this proposal have been considered in the literature in the past years (for a sample, see \cite{Polychronakos:2001mi,Barbon:2001dw,Fradkin:2002qw,Cappelli:2006wa}). 

The holographic duals of the non-commutative field theories can be obtained by performing a series of string dualities to the supergravity backgrounds dual to commutative theories. In addition to deforming the geometry, these dualities introduce a Neveu-Schwarz $B_2$ field, in agreement with the expectations based on the results of \cite{Seiberg:1999vs}.  For the maximally supersymmetric Yang-Mills theory this solution was obtained following these methods in 
\cite{Hashimoto:1999ut,Maldacena:1999mh}. Alternatively, the same background has been obtained  \cite{Li:1999am}  by identifying the field theory metric with the open string metric and using the relations of \cite{Seiberg:1999vs}. 

The so-called  ABJM theory \cite{Aharony:2008ug} is the $(2+1)$-dimensional analogue of the  4d maximally supersymmetric  Yang-Mills theory. This model is a $U(N)\times U(N)$ Chern-Simons gauge theory in 3d with levels $(k,-k)$ coupled to matter fields in the bifundamental representation of the gauge group.  When $N$ and $k$ are large, this model has a dual holographic description in terms of $AdS_4\times {\mathbb{CP}}^3$ geometry with fluxes in Type IIA supergravity.  Moreover, the ABJM theory can be generalized by adding flavors, \ie, fields transforming in the $(N,1)$ and $(1,N)$ representations of the gauge group \cite{Hohenegger:2009as,Gaiotto:2009tk}.  In the gravity dual these flavors correspond to D6-branes wrapping an ${\mathbb{RP}}^3$ inside the internal ${\mathbb{CP}}^3$ manifold and extended along $AdS_4$.

When the number of flavors is small one can study the system in the quenched approximation, in which the D6-branes are considered as  probes in the  $AdS_4\times {\mathbb{CP}}^3$  background. Finding the backreacted supergravity solution, which corresponds to dealing with unquenched flavors on the field theory side, is a difficult task. 
However, in the Veneziano limit \cite{Veneziano:1976wm} in which both the number of colors $N$ and flavors $N_f$  is large but their ratio $N/N_f$ is fixed,  one can obtain a gravity dual of the unquenched flavored model by employing the smearing technique (see \cite{Nunez:2010sf} for a review).  These ABJM flavored geometries were found in \cite{Conde:2011sw} for the massless flavors (at finite temperature in \cite{Jokela:2012dw}) and in \cite{Bea:2013jxa} for the massive flavors.

By turning on an internal worldvolume flux on the probe D6-branes in the smeared flavored backgrounds one can break parity in the $(2+1)$-dimensional gauge theory and realize the quantum Hall effect \cite{Bea:2014yda}. In view of this connection, it is quite natural to study the non-commutative generalization of the ABJM  geometry as a first step to implement holographically the proposal of \cite{Susskind:2001fb} in the ABJM theory. At the level of field theory, the  non-commutative ABJM  model has been recently studied in perturbation theory in \cite{Martin:2017nhg}, whereas its gravity dual has been obtained using string dualities in \cite{imeroni}. Interestingly, it was pointed out later in \cite{Colgain:2016gdj} that this particular TsT-duality will preserve the same amount of supersymmetry as the ambient commutative theory.

In this paper we construct the gravity dual of the NC ABJM theory with unquenched flavor. Our flavors are massive, which means that we have a dimensionful parameter (the quark mass) at our disposal. We have a theory with two different deformation parameters (due to the non-commutativity  and the flavor) and we are interested in exploring how several observables behave when these parameters are varied.

The rest of this paper is organized as follows. In Section \ref{sec:setup} we will briefly review the gravity dual of the commutative unquenched massive ABJM and then describe its non-commuting refinement. We also make several consistency checks. In Sections \ref{sec:point} and \ref{sec:twopoint} we turn to applications and discuss various settings which probe the geometry and the underlying non-commutativity. Here we are primarily interested in discussing the flavor aspects on top of more conventional non-commutative results obtained in similar setups in the literature. Section \ref{sec:EE} will focus on holographic entanglement entropy in the strip geometry. Section \ref{sec:conclusions} contains our conclusions together with a short outlook. Technical details are relegated to several appendices.


\section{Background geometry}\label{sec:setup}

In this section we will detail the background geometry that will be our base in the following sections where we explore physics with different probes. We will start by writing down the essential background material. We will proceed in steps and by first starting reviewing some essentials of the unquenched massive ABJM geometry constructed in \cite{Bea:2013jxa}. We will then move on to applying a TsT-transformation \cite{Lunin:2005jy} on this geometry, which on the field theory side has an interpretation of deforming the commutative ABJM Chern-Simons matter theory into a non-commutative version.  We also discuss several consistency checks that we have performed to demonstrate that the resulting geometry after the TsT transformation is sensible.

\subsection{Commutative unquenched ABJM}

Before jumping into the non-commutative version of the story, let us start by recalling the commutative geometry. In the next subsection we will introduce a deformation which renders the bulk dual of the commutative ABJM to a non-commutative one.

The metric of the ten-dimensional geometry is the following
\be
  ds^2 = h(x)^{-1/2}\left(-dx^2_{0}+dx^2_{1}+dx^2_{2} \right)+h(x)^{1/2}\left[\frac{e^{2g(x)}dx^2}{x^2}+e^{2f(x)}ds^2_{\mathbb{S}^4}+e^{2g(x)}ds^2_{\mathbb{S}_f^2} \right]\label{eq:10dmetric} \ ,
\ee
where the radial coordinate $x$ is such that the boundary is at $x=\infty$ and the deep IR at $x=0$. The point $x=1$ will play a special role as it corresponds to the energy scale for the onset of unquenched flavors. The internal space is a deformed version of the ${\mathbb{CP}}^3$, which we have chosen to present as a squashed ${\mathbb{S}}^2$-bundle over an ${\mathbb{S}}^4$.
The background is also endowed with Ramond-Ramond fluxes and a non-trivial dilaton, but we will postpone introducing them until next subsection. At this point, in addition to certain features of the dilaton, we will be content in understanding how the metric behaves at different radial positions.

First of all, we note that we denote by $x$ the radial coordinate, which is related to the canonical $r$ variable via
\be\label{eq:radialcoordinate}
 e^{g(x)}\frac{dx}{x} = dr \ .
\ee
In general, $g(x)$ is a complicated function of the radial coordinate. The background geometry has a mass scale, which we denote by $r_q$. This is the geometrical distance from the Poincar\'e horizon $r=0$ to the position where the background has a nonzero support from the smeared D6-branes \cite{Bea:2013jxa}. The $r_q$ can be easily extracted from computing the Nambu-Goto action for a stretching string between $x=0$ and the tip of the D6-branes: $r_q = \int_0^1 dx e^{g(x)}/x$, so that the mass of these strings is $m_q=r_q/(2\pi\alpha')$.

As just mentioned, the background in question has smeared D6-branes. This is a shorthand for saying that we have constructed a fully backreacted geometry with $N_f$ D6-branes acting as sources to the Einstein/Maxwell equations in the Veneziano limit. In addition to the mass scale $r_q$ their presence is measured in terms of a deformation parameter
\be
 \hat\epsilon = \frac{3}{4}\frac{N_f}{k} = \frac{3}{4}\frac{N_f}{N}\lambda \ ,
\ee
where we have introduced the 't Hooft coupling $\lambda = N/k$. The parameter $\hat\epsilon$ can take any value between 0 and $\infty$. We also use an another flavor parameter, $b$, which is related to $\hat{\epsilon}$ by
\begin{align}
   b = \frac{16+13\hat{\epsilon}-\sqrt{16+16\hat{\epsilon}+9\hat{\epsilon}^2}}{12+8\hat{\epsilon}}.
\end{align}
The parameter $b$ takes values between $1$ and $5/4$, where the former corresponds to $\hat{\epsilon}=0$.

In the limit $\hat\epsilon=0$, the background reduces to the usual ABJM geometry, and using the original radial coordinate (\ref{eq:radialcoordinate}), the functions appearing the metric read
\bea
 e^{f(r)} & = & r \label{eq:fIR}\\
 e^{g(r)} & = & r \label{eq:gIR}\\
 h(r) & = & \frac{L_{IR}^4}{r^4} \label{eq:hIR} \ .
\eea
Note that we introduced a radius of curvature
\be
 L_{IR}^4 = 2\pi^2\frac{N}{k} \ ,
\ee
and explicitly attach IR as a subscript. With a little bit of algebra, one can then show that the metric in (\ref{eq:10dmetric}) indeed reduces to that of a standard presentation of the ABJM background with a geometry of $AdS_4\times {\mathbb{CP}}^3$ \cite{Bea:2013jxa}.

At generic values of the deformation parameter $\hat\epsilon>0$, the functions $f,g,h$ have complicated radial dependences. However, they continue to asymptote to the IR solutions in (\ref{eq:fIR})-(\ref{eq:hIR}). They also asymptote to the same forms in the UV, albeit with a different radius of curvature, this time denoted by
\be
 L_{UV}^4 = 2\pi^2\frac{N}{k}\sigma(\hat\epsilon)^2 \ , \\
\ee
where $\sigma(\hat\epsilon)$ is the so-called screening function,
\begin{align}
   \sigma = \sqrt{5-4b}(b-2)b \ , \label{eq:screening_func}
\end{align}
which encodes the effects of flavor degrees of freedom. The function $\sigma$ is equal to unity in the limit of no flavors $\hat\epsilon\to 0$ ($b\to 1$) and it monotonically decreases as a function of $\hat\epsilon$ and asymptotes to 0 as $\hat\epsilon\to\infty$ ($b\to 5/4$). This function enters, for example, as a factor of quark-antiquark potential, thus explaining our choice of words as screening the interactions due to flavor degrees of freedom in the close proximity.

Both in the IR and at the UV, the dilaton assumes a constant form. This then means that both limits $x\to 0$ and $x\to\infty$ are asymptotically AdS$_4$ geometries. But since their radii of curvatures are different, the resulting interpolating geometry described by the metric (\ref{eq:10dmetric}) is the gravity dual of an RG flow between two different conformal fixed points.

\subsection{TsT transformed background}\label{sec:TsT}

Let us now continue by presenting a new solution of Type IIA supergravity which is the non-commutative version of the unquenched massive flavored ABJM background of \cite{Bea:2013jxa}, and reviewed in the preceding subsection. The solution is obtained using the engineering of TsT transformations presented in \cite{imeroni}. 

The metric of this solution reads as follows
\be
  ds^2 = h(x)^{-1/2}\left(-dx^2_{0}+M(x) (dx^2_{1}+dx^2_{2} ) \right)+h(x)^{1/2}\left[\frac{e^{2g(x)}dx^2}{x^2}+e^{2f(x)}ds^2_{\mathbb{S}^4}+e^{2g(x)}ds^2_{\mathbb{S}^2_f} \right]\label{nc10dmetric} \ .
\ee
The commutative massive ABJM background has several fluxes turned on, which we mentioned in passing. The TsT procedure, which we simply refer to as a rotation in the following, amounts to adding non-vanishing magnetic-type fluxes as well. The complete set of fluxes are 
\bea
  F_2 & = & \frac{k}{2}\left(E^1\wedge E^2 - \eta(x)\left(\mathcal{S}^\xi\wedge\mathcal{S}^3+\mathcal{S}^1\wedge\mathcal{S}^2     \right) \right) -  \Theta  \frac{e^{g(x)}}{x}K(x) dx_0 \wedge dx \label{ncF2} \\
  F_4 & = & M(x) dx_1 \wedge dx_2 \wedge \left( \frac{e^{g(x)}}{x} K(x) dx_0 \wedge dx \right. \nonumber \\ 
  & & + \left. \Theta \frac{k}{2}  h(x)^{-1} \left(E^1\wedge E^2 - \eta(x)\left(\mathcal{S}^\xi\wedge\mathcal{S}^3+\mathcal{S}^1\wedge\mathcal{S}^2     \right) \right)\right)\label{ncF4} \\
  B_2 & = & - \frac{ \Theta M(x)}{h(x)}    dx_1 \wedge dx_2  \label{ncB2} \\
  H_3 & = & \Theta \frac{M(x)^2}{h(x)^2} h'(x)   dx_1 \wedge dx_2  \wedge dx \label{ncH3} \ .
\eea
To complete the description of the background geometry, we still need to write down the dilaton
\be
  e^{\Phi(x)}=\sqrt{M(x)} e^{\phi(x)} \ ,  \label{ncdil}
\ee
where the function $M(x)$ is defined as
\be
  M(x)=\frac{h(x)}{h(x)+\Theta^2} \label{ncM} \ .
\ee
For more details behind this solution, see Appendix \ref{sec:DetailsBackground}.

Let us make a few remarks on this solution. First, notice that the metric (\ref{nc10dmetric}) is the same as the unrotated version (\ref{eq:10dmetric}), except for the appearance of the function $M(x)$ multiplying the spatial Minkowski directions. This function interpolates monotonically between 1 in the IR, where ABJM theory is recovered, and 0 in the UV, where the spatial dimensions in the $(x_1,x_2)$-plane pinch off, which is standard. Second, there is a NS $B_2$ field that generates a non-vanishing $H_3$. Third, we write the dilaton $\Phi$ in the non-commutative geometry in terms of the unrotated, commutative dilaton $\phi$. We remind the reader that the commutative dilaton $\phi(x)$ is not constant, but only asymptotically in the IR and in the UV, corresponding to the ABJM and massless unquenched ABJM fixed points, respectively. The rest of the functions are the same as in the unrotated case. 

Finally, the TsT transformation amounts to bringing in a new parameter in the theory. This is the angle of the rotation in the T-dual theory, $\Theta$, which in the dual gauge theory corresponds to the non-commutative parameter measuring how much the commutator of the spatial field theory coordinates differs from zero. It is therefore paramount to keep in mind that the theory we are describing has two scale parameters. The first one is the angle $\Theta$ (in units of the curvature radius) and the other one is the mass of the fundamental degrees of freedom that we denoted by $r_q$. Together they will be combined into one dimensionless parameter. This combination can be directly read off from the expression in the function $M(x)$, 
\be
 \Theta \to r_q^2 \Theta \ .
\ee
In the same vein, all the other length scales that appear in the subsequent subsections will be written multiplied by $r_q$ (recall that $r_q$ is an energy scale). Presentations of the results at different amounts of non-commutativity depends on the magnitude {\emph{relative}} to the intrinsic energy scale of the system $r_q$ at a given number of flavors $\hat\epsilon$ in the background. 

In the paper at hand, we are interested in two different effects. First, we wish to understand what happens when we exit the well-understood commutative regime and begin to probe the UV, where non-commutativity should set in and, in particular, what are the novel effects derived from the dynamical quarks in the system. Second, the number of flavors $\hat\epsilon$ is another free parameter in our model that we can dial. It is therefore of great interest to draw lessons of physics interest of how the non-commutative behavior is affected with increasing numbers of unquenched quarks.

\subsection{Consistency checks}

Before starting to discuss applications of the non-commutative background generated above let us lay out several consistency checks that we have performed. The TsT transformation should be viewed as a solution generating technique with which one can obtain a magnetic field in the ten-dimensional geometry without much effort. The solution one ends up with is not, however, a small perturbation of the original solution and one could cast doubt on the sanity of the procedure. For example, the non-commutative directions $(x_1,x_2)$ have a non-trivial factor $M(x)$ multiplying them in the metric (\ref{nc10dmetric}) such that these directions completely collapse at the boundary. This is of course expected behavior in the dual bulk geometries of the non-commutative field theories. Nevertheless, it is still worthwhile to invest effort in understanding the background we are dealing with.

First of all, as detailed in Appendix~\ref{sec:SUSY2} we check that the theory is ${\cal N}=1$ supersymmetric, as long as the same BPS equations as in the unrotated case are satisfied. That is, a solution in the unrotated case is a solution of the rotated one, as the TsT procedure proposes. The set of BPS equations can be combined into one master equation, see \cite{Bea:2013jxa}. Thus, any solution of the master equation is also a solution in the non-commutative case. In particular, the unquenched massive solution we are interested in is a valid solution in the non-commutative sense. Explicitly, the projections imposed on the Killing spinor are
\be
  \Gamma_{47} \epsilon = \Gamma_{89} \epsilon=\Gamma_{56} \epsilon  ~~~~~~~~~~~~  \Gamma_{012}  \epsilon =\sqrt{M} \left( -1+ \frac{\Theta}{\sqrt{h}} \Gamma_{12} \Gamma_{11} \right) \epsilon ~~~~~~~~~~ \Gamma_{3458} \epsilon=- \epsilon \ , \label{killingspinorprojectionssummary}
\ee
where the second projection is the rotated version of the commutative one, while the others are the same.

Second, in Appendix~\ref{sec:kappa} we make another non-trivial verification of the well-behavedness of the background. Namely, we will embed a massive flavor D6-brane in this non-commutative background and by direct computation show that the resulting configuration is kappa symmetric. The embedding of the massive flavor brane in the non-commutative background satisfy kappa symmetry with exactly the same profile for the embedding 
as in the commutative case.

In what follows we will be needing the open string metric \cite{Seiberg:1999vs} which is the effective metric seen by open strings propagating in a background $B_2$-field. The open string metric is defined by
\begin{align}
   {\cal G} = g - B_2 g^{-1} B_2 \ . \label{eq:open_string_metric}
\end{align}
When explicitly evaluated in our geometry, the open string metric is
\bea
   {\cal G} & = & g + \frac{\Theta^2}{h(x)^{3/2}} M(x) (dx_1^2 +dx_2^2) \\
   & = & h(x)^{-1/2} \Bigg( -dx_0^2 + \underbrace{(\frac{\Theta^2}{h(x)}+1) M(x)}_{=M(x)^{-1}M(x)=1}(dx_1^2+dx_2^2) \Bigg) \nonumber\\ 
   & & + h(x)^{1/2} \left( \frac{e^{2 g(x)} dx^2}{x^2} + e^{2f(x)} g_{S^4} + e^{2g(x)}g_{S^2} \right) \\
   & = & g(\Theta=0) \ .
\eea
From the point of view of open strings, non-commutativity is not present in the metric. Even though the open string metric is unaffected by non-commutative effects, non-commutativity still affects computations through the dilaton. In order to make it clear which metric we are referring to, by closed string metric we mean $g$.


\section{Hanging and spinning strings}\label{sec:point}

In this section we will start looking into physical applications. We begin with string probes such that both their endpoints are held at the boundary. The first application considers hanging open strings from the boundary and determining their embedding functional dynamically in the unquenched non-commutative background. The study of this problem sheds light on a gauge invariant observable in the field theory side, namely the Wilson loop. In the subsequent section we will give the open strings a spin, an extreme limit of the rotation corresponding to the study of meson spectrum with high-spin.

\subsection{Wilson loop} \label{wilsonloopsection}

In this section we obtain the expectation value of the Wilson loop and the quark anti-quark $q\bar{q}$ potential for our model. We mostly follow \cite{Alishahiha:1999ci}, while related work can be found in \cite{Bigatti:1999iz,Mateos:2002rx,Maldacena:1999mh}.

In non-commutative quantum field theories dipoles are natural elements \cite{Matusis:2000jf}. A generic feature characteristic of these theories is that the length of the dipoles is related to their transverse momentum. Accordingly, we consider a $q\bar{q}$ configuration with a non-vanishing transverse momentum. Our results verify this generic expectation.

In general, an infinite magnetic field limit imposes restrictions on the charged particle configuration space. For example,  a single charged particle in the presence of a magnetic field describe cyclotron trajectories, in which the radius, velocity, charge, and mass are related. The limit $B\rightarrow \infty$ imposes further restrictions: the cyclotron radius goes to 0, so a single charged particle must remain static. 
Similarly, if we have two charged particles of opposite sign, the limit $B\rightarrow \infty$  imposes also a severe restriction: the distance between the charges is related to the transverse momentum of the dipole \cite{Bigatti:1999iz}. Notice that this is a non-local behavior.
This intuition also underlies non-commutative field theories, which in many cases are related to infinite magnetic field limits.

The holographic prescription for the computation of the Wilson loop was introduced in \cite{Rey:1998ik,Maldacena:1998im}. In that case, the idea is to take a D3-brane out of the stack of D3-branes to infinity and then perform the decoupling limit. A string connecting this D3-brane with the stack of branes will correspond to an infinite massive quark, and the holographic prescription for obtaining the Wilson loop corresponds to computing the string proper length of a $q\bar{q}$ pair living on this D3-brane. In the non-commutative version \cite{Maldacena:1999mh} the idea is to consider a non-vanishing $B_2$ field in the initial D3-brane configuration and take the limit $B_2\rightarrow\infty$ also in the decoupling limit. 
A similar procedure can be carried out in the case of the ABJM theory, where a D2-brane can be taken to infinity in the initial brane construction \cite{Aharony:2008ug}, and also in the presence of a $B_2$ field and flavor branes.

We consider the action of a fundamental open string worldsheet which at the boundary describes a rectangular loop as in \cite{Maldacena:1998im}. The dynamics of these open strings, in the presence of $B_2$, mimic the electric dipoles \cite{SheikhJabbari:1999vm}. In the presence of a $B_2$ field it is given by the Nambu-Goto plus Wess-Zumino action
\be\label{string_action}
S_{string}=-\frac{1}{2 \pi \alpha'} \int d\tau d\sigma \sqrt{-\text{det} \ \hat{g}_2}  + \frac{1}{2 \pi \alpha'} \int \hat{B_2} \ ,
\ee
where $\hat{g}_2$ and $\hat{B}_2$ are the induced metric and $B_2$ field on the worldsheet of the string.
According to the discussion above, the natural configuration of a quark-antiquark pair in our background suggest a string configuration given by $t=\tau$, $x_1=\sigma$, $x_2=v \tau$, $x=x(\sigma)$ with the endpoints located at $x_1=\pm d/2$.\footnote{If one computes the Wilson loop with the configuration of a static string, the endpoints of the string can not be attached to infinity. See, for example, \cite{Maldacena:1999mh}.} We restrict the velocity $v$ to be subluminal as this corresponds to keeping the root square real in (\ref{string_action}).

The induced metric and $B_2$ field are
\bea
d\hat{s}_2^2 & = & h^{-1/2}(M v^2-1)d\tau^2+ h^{-1/2}( M +h\frac{e^{2g}}{x^2}x'^2) d\sigma^2  \\
\hat{B}_2    & = & v \Theta \frac{M}{h} d\tau \wedge d\sigma  \ ,\label{induced_fields}
\eea
Then, the action (\ref{string_action}) reads
\be\label{string_action2}
S_{string}=-\frac{1}{2 \pi \alpha'} \int d\tau d\sigma \left(\sqrt{(1-Mv^2)\left( h^{-1}M+\frac{e^{2g}}{x^2}x'^2\right)}  -v \Theta \frac{M}{h}  \right) \ .
\ee
As the action does not depend on 'time', the 'Hamiltonian' is conserved
\be\label{conserved_quantity}
\frac{M\sqrt{1-Mv^2}}{\sqrt{h(M+h \frac{e^{2g}}{x^2}x'^2)}}-v \Theta \frac{M}{h}=\text{constant}  \ .
\ee
Expanding around the UV we obtain that the constant is $-\frac{v}{\Theta}$. Solving for $x'$,
\be\label{expression_for_xprime}
x'=\pm \frac{x \sqrt{\Theta^2-h v^2}} {v e^g h} \ .
\ee
At the tip of the string, that we denote by $x_*$, the derivative must be zero for the profile to be smooth, imposing ($h_*\equiv h(x_*)$)
\be\label{relationvthetax}
h_* = \frac{\Theta^2}{v^2}  \ .
\ee
Then, we have
\be\label{quark_separation}
d=2 \int_{x_*}^{\infty}\frac{h e^g}{x \sqrt{h_*-h }} dx  \ .
\ee
Expression (\ref{quark_separation}) relates the quark separation $d$ to the tip of the brane $x_*$, and expression  (\ref{relationvthetax}) further relates the location of the tip $x_*$ to the velocity $v$ and the non-commutative parameter $\Theta$. So, we obtain that the transverse velocity $v$ of the dipole is related to its separation $d$, as expected from the discussion above.

The on-shell action is
\be\label{onshell_action}
S_{string}^{on-shell}= -\frac{T}{\pi}\int_{x_*}^{\infty}\frac{\Theta e^g}{x \sqrt{\Theta^2-h v^2 }} dx= -\frac{T}{\pi} \int_{x_*}^{\infty}\frac{\sqrt{h_*} e^g}{x \sqrt{h_*-h  }} dx  \ ,
\ee
where $T=\int d\tau$. This expression is divergent due to the infinite mass of the quarks. In order to remove this infinite mass, we may consider an isolated quark moving with velocity $v$. But according to the discussion above, an isolated charge must remain static in our background. In fact, a straight string is a solution to the equation of motion only if it is static. So, we subtract from (\ref{onshell_action}) the action of two straight static strings and take the large $T$ limit to obtain the $q\bar{q}$ potential energy:
\be\label{eq:renormalized_onshell_action}
 E_{q\bar{q}}= \frac{1}{\pi} \int_{x_*}^{\infty} \frac{e^g}{x} \left( \frac{\sqrt{h_*}}{\sqrt{h_*-h  }}-1 \right) dx -\frac{r_*}{\pi} \ ,
\ee
where $r_*$ is the turning point in the $r$ coordinate: 
\be\label{r_turning_point}
r_* = \int_0^{x_*} \frac{e^g}{x} dx  \ .
\ee

The integrals (\ref{quark_separation}) and (\ref{eq:renormalized_onshell_action}) are the same as in the commutative case \cite{Bea:2013jxa},\footnote{See expressions (8.6) and (8.10) there.} where the 't Hooft coupling $\lambda$ is replaced by its non-commutative counterpart $\lambda \Theta$, according to \cite{Seiberg:1999vs}. This is consistent with the general intuition of the TsT-transformation, which provides the same equations but in a rotated fashion. 

The numerical solutions to (\ref{quark_separation}) and (\ref{eq:renormalized_onshell_action}) were obtained in  \cite{Bea:2013jxa}, see Fig. 7 there.\footnote{In that figure, in the horizontal axes label, we must replace $\lambda$ by $\lambda\Theta$ in the non-commutative case.} The result is that the $q\bar{q}$ energy smoothly interpolates between the $1/d$ behaviors at the IR and the UV.
At the endpoints of the flow, expressions (\ref{quark_separation}) and (\ref{eq:renormalized_onshell_action}) can be computed exactly. At the UV we obtain:
\be\label{eq:renormalized_onshell_actionABJMUV}
E_{q\bar{q}}^{UV}=- \frac{4\pi^3 \sqrt{2\lambda  \Theta}}{\Gamma(1/4)^4} \ \sigma \ \frac{1}{d} ~, ~~~~~~~~~~~~~~~~ r_q \ d \rightarrow 0 \ ,
\ee
where $\sigma$ is the screening function defined in (\ref{eq:screening_func}). In the IR:
\be\label{eq:renormalized_onshell_actionABJMIR}
 E_{q\bar{q}}^{IR}=- \frac{4\pi^3 \sqrt{2\lambda \Theta}}{\Gamma(1/4)^4} \ \frac{1}{d} ~, ~~~~~~~~~~~~~~~~~~~~ r_q \ d \rightarrow \infty \ .
\ee
This last expression is the same as for the pure ABJM theory.
For more details of the solutions to (\ref{quark_separation}) and (\ref{eq:renormalized_onshell_action})  see Section 8 and Appendix D of \cite{Bea:2013jxa}.
 
The subluminal condition $v<1$ together with (\ref{relationvthetax}) and (\ref{quark_separation}) imply that there is a minimal $q\bar{q}$ distance given by the non-commutative parameter $d_{min} \sim 1/\sqrt{\Theta}$. This is consistent with the spatial version of the uncertainty principle, characteristic of non-commutative field theories. Moreover, notice that for large dipole lengths, when the quarks are very separated from each other, we obtain that the velocity goes to zero, consistently with the fact that isolated quarks must remain static.
 
In addition to the Wilson loops, other natural observables in non-commutative field theories are the Wilson lines. In these theories, the Wilson lines are gauge invariant as long as the dipole length and its transverse momentum verify a specific relation \cite{Matusis:2000jf}.  We leave the study of these observables for the future.

\subsection{Rotating string in the non-commutative plane}

In this section we study the spectrum of mesons with large spin. In the limit of large spin, we may work in the semiclassical approximation where string fluctuations are neglected. The string rotates in the non-commutative plane $(x_1,x_2)$ and extends inwards into the bulk along the holographic coordinate. In contrast to the previous prescription in the computation of the Wilson loop holographically, here we are forced to introduce an additional D2-brane close to the boundary where the string endpoints at attached \cite{Arean:2005ar}, {\emph{i.e.}}, its endpoints lie on a D2-brane which spans the non-commutative plane and is placed at $r=r_{D2}$ along the holographic direction. Notice that such a domain-wall D2-brane will remain at $r_{D2}$ as it is compatible with the D2-brane fluxes of the background. 

We will now find solutions to equations of motion for the rotating string, extract energies and angular momenta of various rotating configurations and study how these quantities are correlated.
The string worldsheet embedding is
\begin{align}
  x_0 &= \tau \\
  x_1 &= \rho(\sigma) \cos(\omega\tau) \\
  x_2 &= \rho(\sigma) \sin(\omega\tau) \\
  r   &= 1/z(\sigma) \ ,
\end{align}
where $\omega$ is a constant angular frequency. Pulling back to the worldsheet the metric and magnetic field are
\begin{gather}
  \hat g_2 = \frac{1}{\sqrt{h}} (-1+\omega^2\rho(\sigma)^2 M) d\tau^2 + \sqrt{h} \left( \frac{z'(\sigma)^2}{z(\sigma)^4} + \frac{\rho'(\sigma)^2 M}{h} \right) d\sigma^2 \\
  \hat B_2 = \frac{\Theta\omega\rho(\sigma)\rho'(\sigma) M}{h} d\tau\wedge d\sigma \ .
\end{gather}
The dynamics of the semiclassical string are governed by the Nambu-Goto action (\ref{string_action}).
By using (\ref{string_action}) together with expressions for $g_2$ and $B_2$ we obtain the Lagrangian
\begin{align}
  \mathcal{L} = \frac{1}{2\pi\alpha'} \left(-\frac{\sqrt{(\omega^2\rho^2 M-1)(h z'^2+M z^4 \rho'^2)}}{ \sqrt{h} z^2} + \frac{\Theta\omega\rho\rho' M}{h}\right)\ . \label{rotatinglagrangian}
\end{align}
The Lagrangian (\ref{rotatinglagrangian}) is cyclic in $t$ and $\theta$. The corresponding conserved quantities are
\begin{align}
  E &= \dot{\theta} \frac{\partial L}{\partial \dot{\theta}} - L = \frac{1}{2\pi\alpha'} \int d\sigma \frac{1}{\sqrt{h}z^2}\frac{\sqrt{h z'^2+M z^4\rho'^2}}{\sqrt{1-\omega^2\rho^2 M}} \label{rotatingenergy} \\
  J &= \frac{\partial L}{\partial \dot{\theta}} = \frac{1}{2\pi\alpha'} \int d\sigma \left( \frac{\omega\rho^2 M}{\sqrt{h}z^2} \frac{\sqrt{h z'^2+M z^4\rho'^2}}{\sqrt{1-\omega^2\rho^2 M}} + \frac{\Theta\rho\rho' M}{h}\right) \ . \label{rotatingangularmomentum}
\end{align}
The last term in $J$ is the Kalb-Ramond term. By $J_1$ we will denote $J$ with this term subtracted off. The Kalb-Ramond term is important in interpreting our results and whether or not we choose to include it in our definition of the angular momentum has a drastic effect on the outcome of our computation.

The profile of the rotating string is determined by the equations of motion derived from the Lagrangian (\ref{rotatinglagrangian}). For these equations to yield actually extremal configurations of the action, we must make sure that suitable boundary conditions on the D2-brane are respected. The boundary conditions must be such that
\begin{align}
  \left.\frac{\partial L}{\partial z'}\delta z'\right|_{\partial\Sigma} = \left.\frac{\partial L}{\partial \rho'}\delta\rho \right|_{\partial\Sigma} = 0 \ .
\end{align}
For $z(\sigma)$ the boundary conditions hold automatically since the variation $\delta z|_{\partial\Sigma}=0$ by the fact that the $z$-coordinates of the string endpoints are fixed. However, since $\delta\rho|_{\partial\Sigma}$ is arbitrary on the brane, we must impose
\begin{align}
  \left.\frac{\partial L}{\partial \rho'}\right|_{\partial\Sigma} = 0 \ .
\end{align}
After inserting (\ref{rotatinglagrangian}), this equation becomes
\begin{align}
  \frac{\Theta\omega M}{h}\rho - \frac{M z^2}{\sqrt{h}} \frac{\sqrt{1-\omega^2\rho^2 M}}{\sqrt{z'^2 h+M z^4}}\rho' = 0 \ . \label{stationarycondition}
\end{align}
Since the coefficients of $\rho$ and $\rho'$ are non-negative, we deduce that if the string endpoints have $\rho$ coordinates $\rho_0>0$ and $\rho_1<0$, the signs of $\rho'$ at these endpoints must be
\begin{align}
  \rho'|_{\rho_0} \geq 0,\qquad \rho'|_{\rho_1} \leq 0 \ ,
\end{align}
respectively. In the commutative case $\Theta=0$, eq (\ref{stationarycondition}) imposes $\rho'|_{\partial\Sigma} = 0$, that is, the string ends on the D2-brane orthogonally. With a finite non-commutativity, the string configuration is tilted. We can use (\ref{stationarycondition}) to solve for $\partial z/\partial\rho$ at the endpoints. The result is
\begin{align}
  \left.\frac{dz}{d\rho}\right|_{\rho_0} = -\frac{z_{D2}^2\sqrt{1-\omega^2\rho_0^2}}{\Theta\omega\rho_0},\qquad \left.\frac{dz}{d\rho}\right|_{\rho_1} = \frac{z_{D2}^2\sqrt{1-\omega^2\rho_1^2}}{\Theta\omega\rho_1}\ . \label{rotatingboundarycondition}
\end{align}
It is clear that this boundary condition is independent of flavors. It is also apparent that the angular velocity $\omega$ is bounded by the endpoint $\rho$-coordinate
\begin{align}
  \omega \leq \frac{1}{\rho_0} \label{omegabound}
\end{align}
and similarly for the other endpoint. It turns out that physical string configurations are those where both string endpoints are at equal distance from the origin $\rho=0$, that is $\rho_1=-\rho_0$. If we denote $\rho_0-\rho_1=\bar{\rho}$, (\ref{omegabound}) implies that in the fast spinning limit the relation between the separation of quarks and the angular velocity must be $\bar{\rho} \sim \omega^{-1}$.

\begin{figure}
  \centering
  \includegraphics[width=0.8\textwidth]{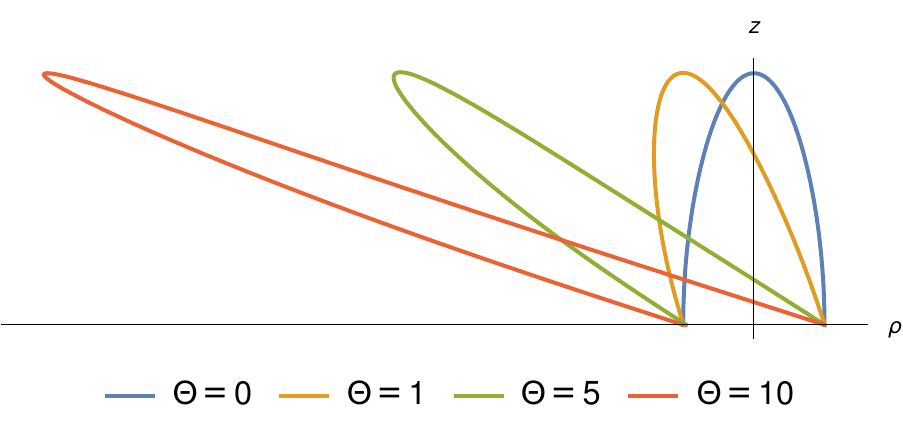}
  \caption{Profiles for rotating strings with various values of $\Theta$ and fixed other parameters. As non-commutativity increases the strings become more tilted. The $z$-coordinate of the deepest point in the bulk is insensitive to non-commutativity.}
  \label{rotatingstringsfig}
\end{figure}

The equations of motion derived from (\ref{rotatinglagrangian}) are messy and not illuminating to quote here. To solve the equations we make the choice of parametrization $\sigma=\rho$. The resulting equation of motion is second order, non-linear, and very complicated. We solve this equation numerically with boundary conditions $z(\rho_0)=z(\rho_1)=z_{D2}$ with $z'(\rho_0)$ and $z'(\rho_1)$ given by (\ref{rotatingboundarycondition}). It is clear that the $\rho=\sigma$ parametrization cannot be viable for the whole length of the string since (\ref{rotatingboundarycondition}) demands that the string must tilt. In practice we impose our boundary conditions on $\rho=\rho_0$ and integrate toward $\rho=\rho_1$ until the string starts to turn back towards the D-brane. At this point we change parametrization to $\sigma=z$ and integrate until the string hits the brane. We perform this repeatedly for different choices of $\omega$ to scan for a configuration such that string boundary conditions are also fulfilled at $\rho=\rho_1$. A few of these configurations are presented in Fig.~\ref{rotatingstringsfig}.

\begin{figure}
  \centering
  \includegraphics[width=0.6\textwidth]{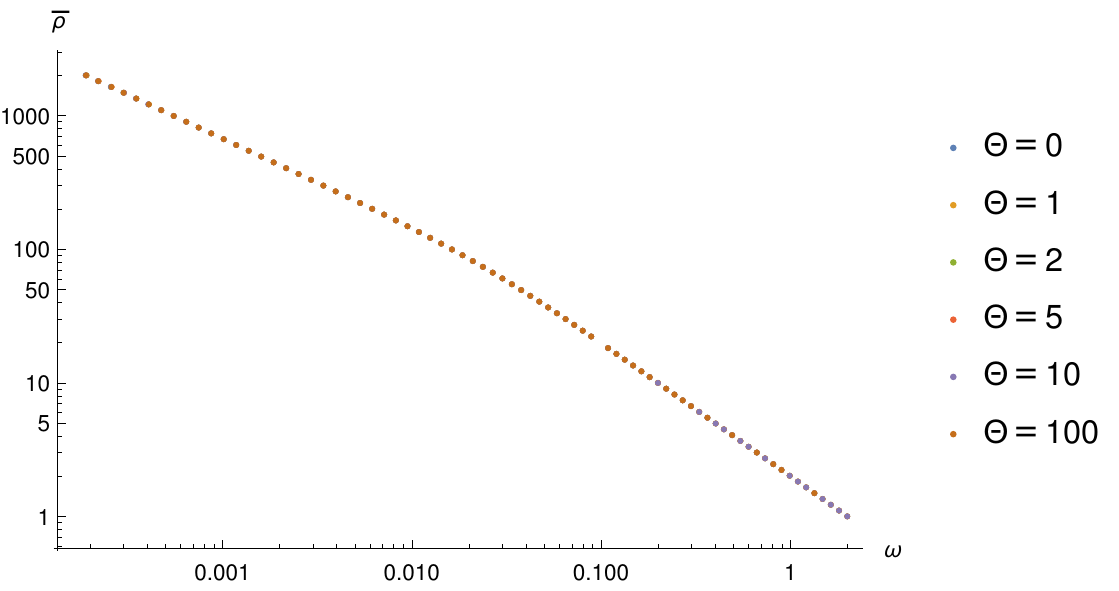}
  \caption{Quark separation as a function of angular velocity with no flavor in the background. Notice the two scaling regimes in (\ref{exponent1}) and (\ref{exponent2}).}
  \label{rhobar_vs_omega_fig}
\end{figure}

We first present results without flavors and later discuss how flavor effects modify the flavorless results. Recall, that in the flavorless case the metric function $h(r)$ is given in (\ref{eq:hIR}).
We have analyzed many string configurations to study correlations between $\omega$, $\bar\rho$, $E$, and $J$. We found that short quark separation $\bar\rho\to 0$ corresponds to fast rotation $\omega\to\infty$ and large quark separation $\bar\rho\to\infty$ corresponds to slow rotation $\omega\to 0$. We have plotted $\bar\rho(\omega)$ in Fig.~\ref{rhobar_vs_omega_fig}.
From the figure it is clear that there are two regimes with different power-law behavior with a reasonably sharp knee separating the two regimes. The power-law exponents are
\begin{align}
  \bar\rho &\sim \omega^{-2/3}, \qquad \text{for } \omega\to 0 \label{exponent1} \\
  \bar\rho &\sim \omega^{-1}, \qquad \;\;\: \text{for }  \omega\to\infty \ . \label{exponent2}
\end{align}

\begin{figure}
  \begin{minipage}{\textwidth}
    \begin{minipage}{0.32\textwidth}
      \includegraphics[width=\textwidth]{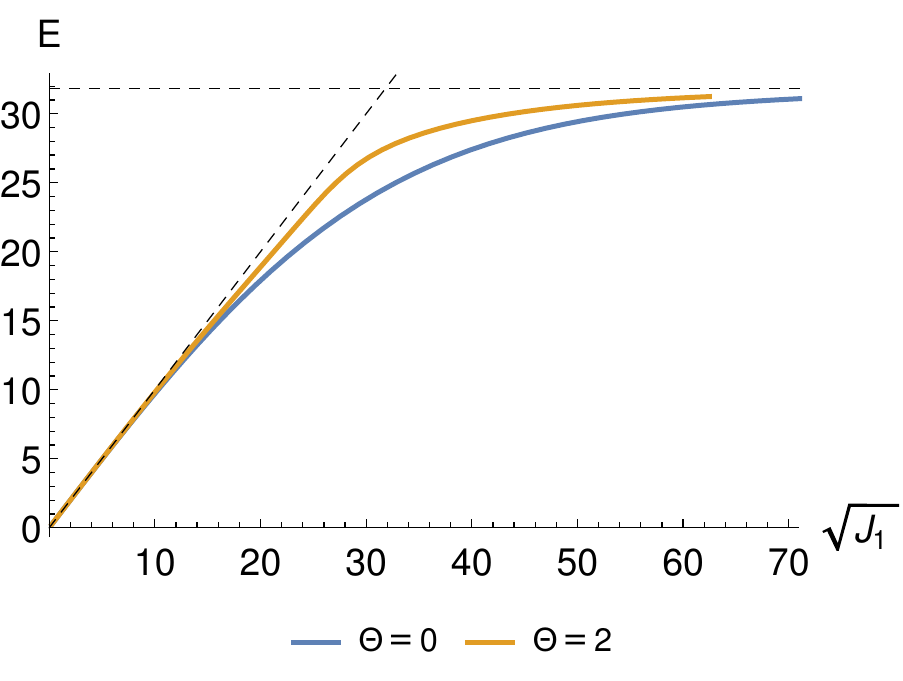}
    \end{minipage}
    \begin{minipage}{0.32\textwidth}
      \includegraphics[width=\textwidth]{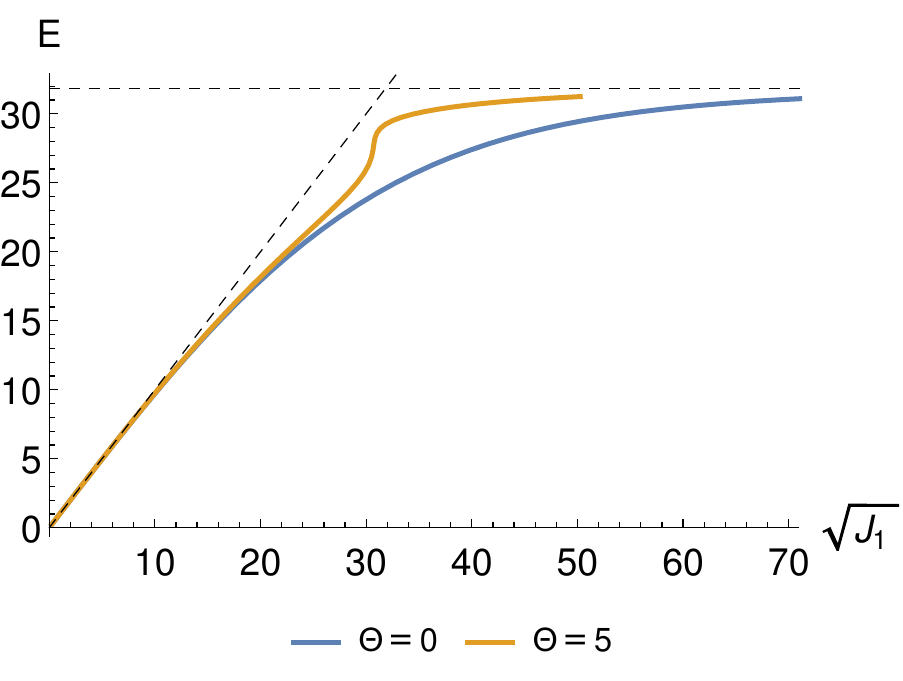}
    \end{minipage}
    \begin{minipage}{0.32\textwidth}
      \includegraphics[width=\textwidth]{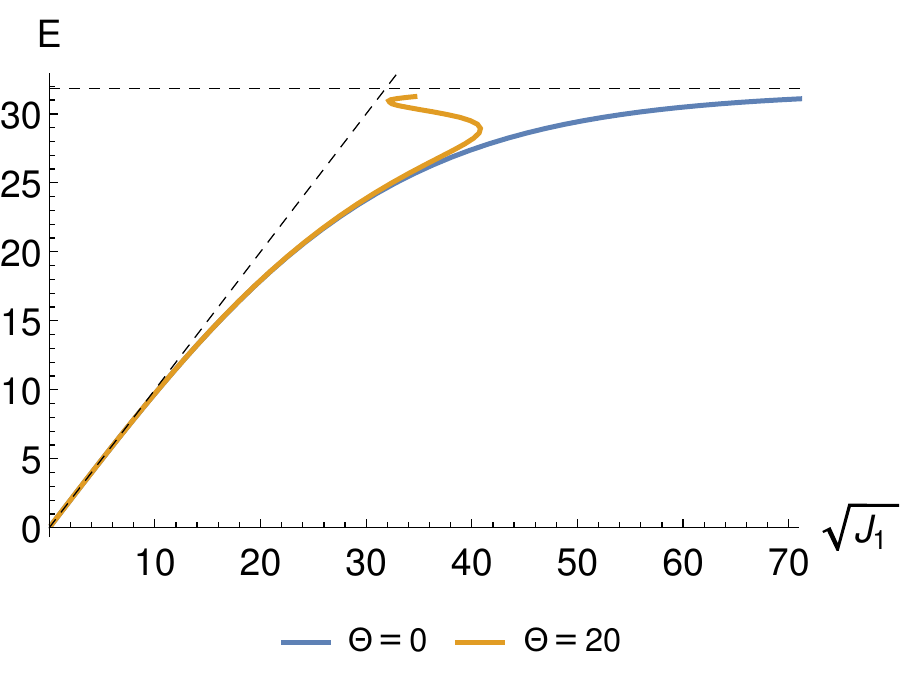}
    \end{minipage}
  \end{minipage}
  \caption{Energy of a rotating string as a function of its angular momentum without added flavor. Meson energies follow Regge trajectories for small quark separation which corresponds to small $J_1$. For large quark separation the energy saturates to the long distance meson energy (dashed line).}
  \label{E_vs_sqrtJ_fig}
\end{figure}

When the extremal string profiles are known, it is straightforward to use (\ref{rotatingenergy}) and (\ref{rotatingangularmomentum}) to compute corresponding energies and angular momenta. Fig.~\ref{E_vs_sqrtJ_fig} shows the relation between energy and angular momentum for multiple values of the non-commutative parameter.
The energy of the string configuration clearly follows a Regge trajectory for small $J_1$ regardless of non-commutativity or flavor (shown later). Compared to commutative theory the effect of turning on non-commutativity is first to increase meson energy for a fixed angular momentum. After the non-commutativity parameter $\Theta$ increases over some critical value, the energy starts to decrease. When $\Theta$ is turned to a very large value, commutative theory is restored as is expected on general grounds in the non-commutative field theories \cite{Seiberg:1999vs}. The fact that the energies are not sensitive to non-commutativity is expected as this corresponds to large quark separation and in such a case the effects of non-commutativity should disappear. Short strings, on the other hand, are not affected much by the background, leading to insensitivity to $\Theta$  in the small $J_1$ regime; see Fig.~\ref{E_vs_sqrtJ_fig}.

Were we to redo Fig.~\ref{E_vs_sqrtJ_fig} with $J$ instead of $J_1$ by adding the contributions coming from the Kalb-Ramond term all the novel features of non-commutativity would disappear as noted in \cite{Haque:2009hz}. The complete angular momentum $J$ is insensitive to non-commutativity whereas the ``physical'' angular momentum $J_1$ is sensitive to it. 

\subsubsection{Regge slope}
In the large angular velocity and large non-commutativity strings stretch and tilt wildly as is illustrated in Fig.~\ref{rotatingstringsfig}. In this limit, rotating strings can be approximated by two straight lines making the analytic computation of $E$, $J_1$, and the Regge slope possible. Let $\rho_*$ denote the $\rho$-coordinate of the string turning point. The value of $\rho_*$ is determined as the place where $d^2 z/d^2\rho$ diverges. By examining the equation of motion for $z(\rho)$, this is seen to happen when
\begin{align}
  1 - M(z) \rho_*^2 \omega^2 = 0\ .
\end{align}
In the limit we are working in, $z$ is always very close to $z_{D2}$, so we approximate $z\sim z_{D2}$. The turning point is
\begin{align}
  \rho_* = \frac{1}{\sqrt{M(z_{D2})} \omega} \approx \frac{\Theta}{\sqrt{h(z_{D2})}\omega} \ .
\end{align}
In the last step we have taken the large $\Theta$ limit. The $z$-coordinate of the turning point is now given by
\begin{align}
  z_* \approx \left. \frac{d z}{d\rho}\right|_{\rho=\rho_0} \rho_* \approx \frac{z_{D2}^2 \sqrt{1-\omega^2 \rho_0^2}}{\sqrt{h(z_{D2})}\omega^2\rho_0}\ .
\end{align}
This formula agrees with our observation that the $z$-coordinate at which the string turns around is quite insensitive to non-commutativity in general.

By approximating $z\sim z_{D2}$, the formulas (\ref{rotatingenergy}) and (\ref{rotatingangularmomentum}) yield
\begin{align}
  E &\approx \frac{1}{\pi\alpha'} \sqrt{\frac{M(z_{D2})}{h(z_{D2})}} \int_0^{\rho_*} \frac{d\rho}{\sqrt{1-\omega^2\rho^2 M(z_{D2})}} = \frac{1}{2\alpha'\omega\sqrt{h(z_{D2})}} \\
  J_1 &\approx \frac{1}{\pi\alpha'} \frac{M(z_{D2})^{3/2}}{\sqrt{h(z_{D2})}} \int_0^{\rho_*} \frac{\rho^2 d\rho}{\sqrt{1-\omega^2\rho^2 M(z_{D2})}} = \frac{1}{4\pi\alpha'\omega^2\sqrt{h(z_{D2})}}\ .
\end{align}
Thus in the high angular velocity regime $E\sim\sqrt{J_1}$. The Regge slope is
\begin{align}
  \alpha'_{eff} = \frac{J_1}{E^2} = \sqrt{h(z_{D2})} \alpha'.
\end{align}
The Regge trajectories are not affected by non-commutativity but they are affected by flavor effects encoded in $h(z_{D2})$.

\subsubsection{Flavor effects}
We can redo the our analysis in the presence of flavors in the background. Everything else carries to flavored case unchanged except the metric function $h(r)$ now interpolates between two $AdS_4$-like behaviors between the IR and UV.

\begin{figure}
  \centering
  \includegraphics[width=0.8\textwidth]{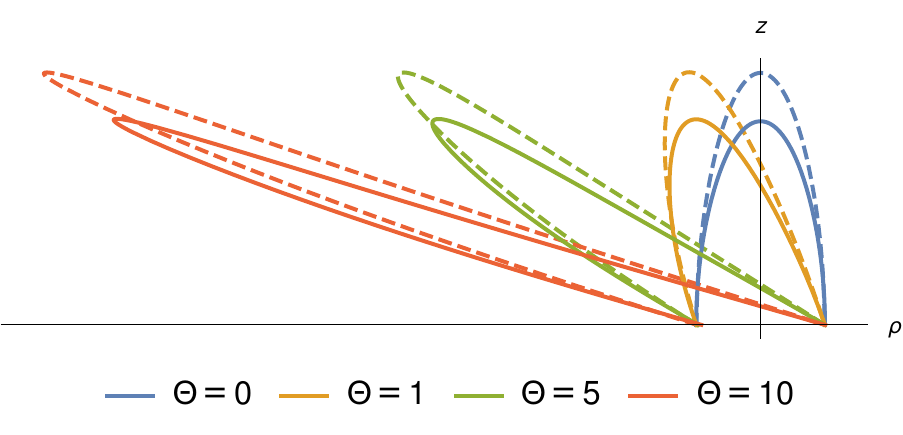}
   \caption{String embeddings for various values of $\Theta$. Dashed curved correspond to the flavored background ($\hat{\epsilon}=1$) and solid lines represent the flavorless case.}
  \label{rotatingstringsflavoredfig}
\end{figure}

\begin{figure}
\begin{minipage}{\textwidth}
  \begin{minipage}{0.32\textwidth}
    \includegraphics[width=\textwidth]{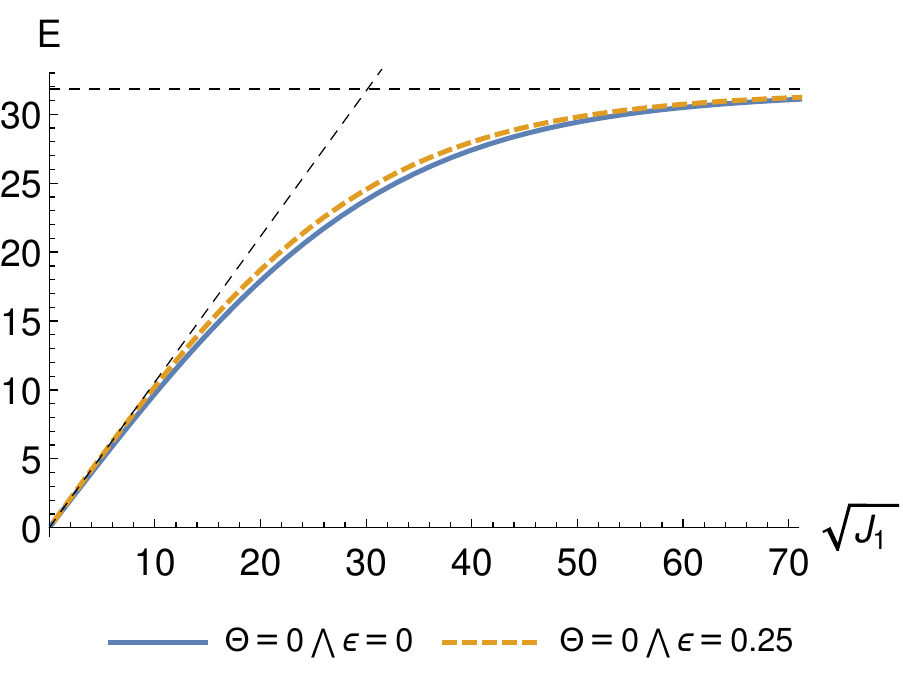}
  \end{minipage}
  \begin{minipage}{0.32\textwidth}
    \includegraphics[width=\textwidth]{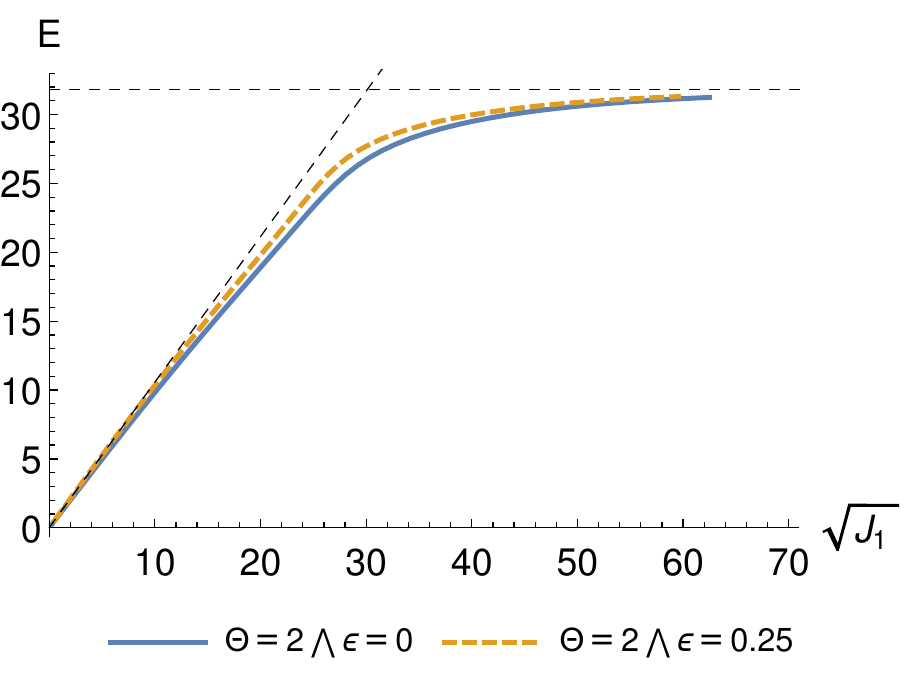}
  \end{minipage}
  \begin{minipage}{0.32\textwidth}
    \includegraphics[width=\textwidth]{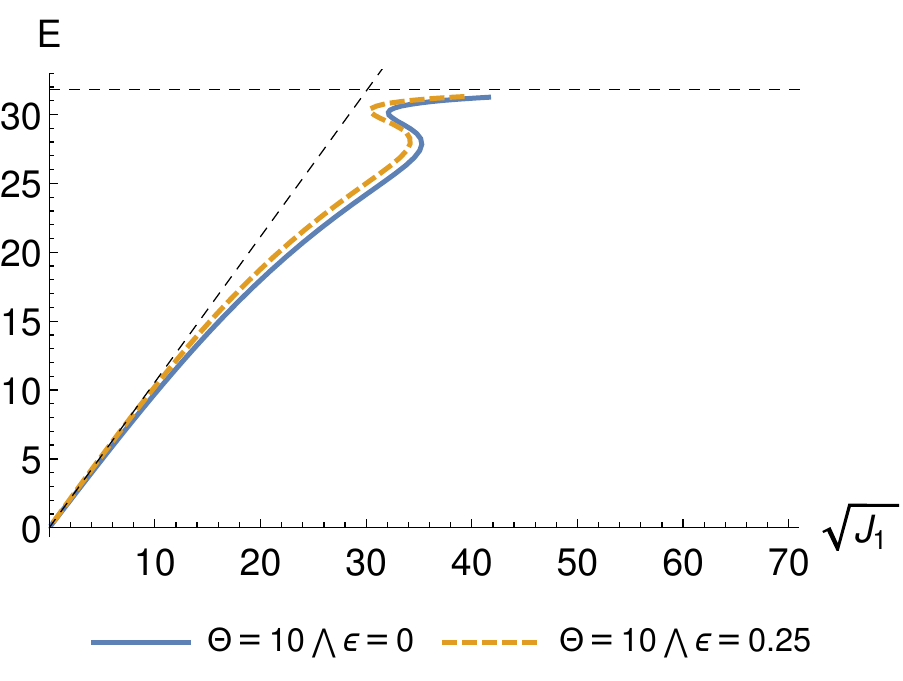}
  \end{minipage}
\end{minipage}
   \caption{$E(\sqrt{J_1})$ for flavorless and flavored ($\hat{\epsilon}=1$) cases for $\Theta=0$ (left), $\Theta=2$ (middle) and $\Theta=10$ (right) cases. Black dashed curves denote the Regge trajectory and the limiting value of energy. The effect of turning on flavor in the background is to increase meson energies for fixed angular momentum.}
\label{reggecomparison}
\end{figure}

When turning on flavors in the background, power law exponents (\ref{exponent1}) and (\ref{exponent2}) characterizing the relation of quark separation to angular velocity remain unchanged. On the other hand, the relation between the meson energy $E$ and angular momentum $J_1$  is affected. The effect is illustrated in figure \ref{reggecomparison}. Turning on flavor increases meson energy for a fixed value of angular momentum. This can be attributed to the dissipation effect. By keeping the angular frequency fixed, it is clear that it takes more energy to sustain this frequency in the soup of unquenched flavors.

\section{Two-point function}\label{sec:twopoint}

Now we turn to the two-point function of massive operators in the non-commutative background. As we are interested only in the very massive operators, we can make use of semiclassical approach and be content by computing the geodesics in the bulk geometry, connecting the operators on the boundary theory at different spatial positions. As emphasized in \cite{Landsteiner:2007bd}, we will use the open string metric (\ref{eq:open_string_metric}) to measure the geodesic length.

We choose to study the equal time two-point function of operators separated in the $x_1$-direction. First we must find the geodesic length which we will minimize. Symmetries of the system imply that the geodesic is of the form $x=x(x_1)$. The Einstein frame metric pulled back to the geodesic is
\begin{align}
   \hat g_1 = e^{-\Phi(x)/4} h(x)^{-1/2} (1+G(x) x'^2) dx_1^2 \ ,
\end{align}
where we have defined $G(x)=h(x)e^{2g(x)}/x^2$. Here it can be seen that, as stated previously, non-commutative effects enter the calculation through the dilaton $e^{\Phi(x)}$ even though the open string metric does not depend on $\Theta$. The length of the geodesic is $\mathcal{L} = \int \sqrt{\det \hat g_1}$, meaning that we should minimize
\begin{align}
   \mathcal{L} = \int e^{-\Phi(x)/4} h(x)^{-1/4} \sqrt{1+G(x) x'^2} dx_1 := \int L(x,x') dx_1 \label{eq:geodesic} \ .
\end{align}
The integrand does not depend on $x_1$, meaning that there is a conserved energy
\begin{align}
   x' \frac{\partial L}{\partial x'} - L = -\frac{e^{-\Phi(x)/4}}{h(x)^{1/4} \sqrt{1+G(x) x'^2}} = - \frac{e^{-\Phi(x)/4}}{h(x_*)^{1/4}} \ .
\end{align}
In the last step we have evaluated the conserved energy at the turning point, denoted by $x=x_*$, where $x'(x_*)=0$. The resulting equation can be used to solve for $x'$. Thus, the slope $x'$ along the geodesic is
\begin{align}
   x' &= \pm \frac{1}{ \sqrt{G(x)}} \sqrt{e^{\frac{1}{2}(\Phi(x_*)-\Phi(x))} \left( \frac{h(x_*)}{h(x)} \right)^{1/2} - 1} \\
   &= \pm \frac{1}{ \sqrt{G(x)}} \sqrt{e^{\frac{1}{2}(\phi(x_*)-\phi(x))} \left( \frac{M(x_*)}{M(x)} \right)^{1/4} \left( \frac{h(x_*)}{h(x)} \right)^{1/2} - 1} \label{eq:geodesic_slope} \ ,
\end{align}
where in the last step we substituted the dilaton, $e^{\Phi(x)} = \sqrt{M(x)} e^{\phi(x)}$. The slope $x'$ can also be used to find the separation $l=\int dx_1=\int dx/x'$ between the two operators, as a function of the turning point
\begin{align}
   l = 2 \int_{x_*}^\infty dx \frac{ \sqrt{G(x)}}{ \sqrt{e^{ \frac{1}{2}(\phi(x_*)-\phi(x))} \left( \frac{M(x_*)}{M(x)} \right)^{1/4} \left( \frac{h(x_*)}{h(x)} \right)^{1/2} -1}} \ . \label{eq:geod_separation}
\end{align}
The length of the geodesic connecting our boundary operators is then given by substituting (\ref{eq:geodesic_slope}) into (\ref{eq:geodesic}). Explicitly,
\begin{align}
   \mathcal{L} = 2 \int_{x_*}^\infty dx \frac{M(x)^{-1/8} e^{-\phi(x)/4} h(x)^{-1/4} \sqrt{G(x)}}{ \sqrt{1-e^{\frac{1}{2}(\phi(x)-\phi(x_*))} \left( \frac{M(x)}{M(x_*)} \right)^{1/4} \left( \frac{h(x)}{h(x_*)} \right)^{1/2}}} \ .
\end{align}
The geodesic length is divergent and must be regularized because the boundary is infinitely far away. We regularize by subtracting the near boundary divergence. The form of this divergence can be found by using the asymptotic expressions of background functions found in Appendix~\ref{app:asymptotics}
The integrand of $\mathcal{L}$ behaves in UV as 
\begin{align}
   \chi(b) \frac{N^{1/4}}{\lambda^{1/16}} \left( \frac{r_q^2 \Theta}{\sqrt\lambda} \right)^{1/4} x^{-1+\frac{1}{2b}} \ ,
\end{align}
where we have defined for convenience
\begin{align}
   \chi(b) = 2^{13/16} \pi^{1/8} b^{-1} \sqrt\kappa \sigma^{1/8} \left( \frac{(2-b)b(1+\hat\epsilon)}{3-2b} \right)^{1/4}
\end{align}
which is only a function of flavor $b$. Notice that the correlator will not be of the CFT form because this divergence is not logarithmic. We define the regularized geodesic length by
\begin{align}
   \mathcal{L}^{reg} = \mathcal{L} - \chi(b) \frac{N^{1/4}}{\lambda^{1/16}} \left( \frac{r_q^2 \Theta}{\sqrt\lambda} \right)^{1/4} \left( \int_{x_*}^\infty x^{-1+\frac{1}{2b}} dx + 2b x_*^\frac{1}{2b} \right) \label{eq:geod_length_reg} \ .
\end{align}
The effect of the subtraction is to remove the asymptotic $x^{-1+\frac{1}{2b}}$-behavior from the integrand, making the integral convergent. The last term is the lower limit of the subtracted integral.

\subsection{UV-limit}
Let us now compute the regularized geodesic length in the UV-limit using (\ref{eq:geod_length_reg}) and asymptotic expansions for background functions
\begin{align}
   \mathcal{L}^{reg} &\approx \chi(b) \frac{N^{1/4}}{\lambda^{1/16}} \left( \frac{r_q^2 \Theta}{\sqrt\lambda} \right)^{1/4} x_*^\frac{1}{2b} \left[ \int_1^\infty z^{-1+\frac{1}{2b}} \left( \frac{z^\frac{3}{2b}}{\sqrt{z^\frac{3}{b}-1}} -1 \right) dz - 2b \right]\\
   &= -2b\sqrt\pi \frac{\Gamma\left( \frac{5}{6} \right)}{\Gamma\left( \frac{1}{3} \right)} \frac{N^{1/4}}{\lambda^{1/16}} \left( \frac{r_q^2 \Theta}{\sqrt\lambda} \right)^{1/4} x_*^\frac{1}{2b} \ . \label{eq:geod_length_reg_UV}
\end{align}
Separation $l$ in (\ref{eq:geod_separation}) can also be evaluated in the UV region by using asymptotic expansions 
\begin{align}
   \frac{r_q l}{\sqrt\lambda} &\approx \frac{2\sqrt{2}\pi\sigma}{b\kappa^2} \int_{x_*}^\infty \frac{x^{-1-1/b}}{\sqrt{\left(\frac{x}{x_*}\right)^{3/b}-1}} = \frac{2\sqrt{2}\pi\sigma}{b\kappa^2} x_*^{-1/b} \int_1^\infty dz \frac{z^{-1-1/b}}{\sqrt{z^{3/b}-1}} \\
   &= \frac{2\sqrt{2}\pi^{3/2}\sigma}{b\kappa^2} \frac{\Gamma\left(\frac{5}{6}\right)}{\Gamma\left(\frac{1}{3}\right)} x_*^{-1/b} \ .
\end{align}
This can be solved for $x_*(l)$ and substituted into (\ref{eq:geod_length_reg_UV}) to obtain the length of the geodesic in terms of the separation $l$. We find
\begin{align}
   \mathcal{L}^{reg} = - \frac{2^{1+3/4} b \pi^{5/4}\sqrt\sigma}{\kappa} \chi(b) \left(\frac{\Gamma\left(\frac{5}{6}\right)}{\Gamma\left(\frac{1}{3}\right)}\right)^{3/2} \frac{N^{1/4}}{\lambda^{1/16}} \left( \frac{r_q^2 \Theta}{\sqrt\lambda} \right)^{1/4} \sqrt{\frac{\sqrt\lambda}{r_q l}} \label{eq:LrefUV} \ .
\end{align}
The two-point function in the semi-classical approximation is thus
\begin{align}
   \langle \mathcal{O}(x)\mathcal{O}(y) \rangle_{UV} \approx e^{-m{\cal L}} = \exp \left[ m C \frac{N^{1/4}}{\lambda^{1/16}} \left( \frac{r_q^2 \Theta}{\sqrt\lambda} \right)^{1/4} \sqrt{ \frac{\sqrt\lambda}{r_q l}} \right] \ ,
\end{align}
where $C$ is a positive, flavor dependent constant that can be read off from (\ref{eq:LrefUV}) and $m$ is the mass of the dual bulk field.

\subsection{Subleading UV-term}
After a tedious computation in parallel to the one presented in \cite{Bea:2013jxa}, the leading and subleading term turns out to be of the form
\begin{align}
   \frac{\mathcal{L}^{reg}}{\frac{N^{1/4}}{\lambda^{1/16}} \left( \frac{r_q^2 \Theta}{\sqrt\lambda} \right)^{1/4}} = C \left(\frac{\sqrt\lambda}{r_q l}\right)^{1/2} + D \left(\frac{r_q l}{\sqrt\lambda}\right)^{2b-1/2} \ ,
\end{align}
where $D$ is a constant depending on $b$ and $r_q \Theta/\sqrt\lambda$.

\subsection{IR-limit}
We split the integration in $l$ to an IR and non-IR part
\begin{align}
   l = 2 \left(\int_{x_*}^{x_a} + \int_{x_a}^\infty\right)\frac{ \sqrt{G(x)} dx}{ \sqrt{e^{ \frac{1}{2}(\phi(x_*)-\phi(x))} \left( \frac{M(x_*)}{M(x)} \right)^{1/4} \left( \frac{h(x_*)}{h(x)} \right)^{1/2} -1}} = l_{IR} + l_{UV} \ .
\end{align}
In the first integral we expand the functions in IR. In the second integral we assume convergence, which is reasonable since we already know that this integral is convergent in the UV. Since in the IR, $h\sim x^{-4}$ and $M=h/(h+\Theta^2)\sim 1$, the IR part must agree with the ABJM result which is \cite{Bea:2013jxa}
\begin{align}
   l_{IR} = 2 L_{IR}^2 \frac{c}{\gamma} \frac{1}{x_*} \ .
\end{align}
Since $x_*\to 0$, we can approximate $l_{UV}$ by
\begin{align}
   l_{UV} &= 2 \int_{x_a}^\infty dx\frac{ \sqrt{G(x)}}{ \sqrt{e^{ \frac{1}{2}(\phi(x_*)-\phi(x))} \left( \frac{M(x_*)}{M(x)} \right)^{1/4} \left( \frac{h(x_*)}{h(x)} \right)^{1/2}-1}} \\
   &= 2 e^{-\frac{\phi(x_*)}{4}} M(x_*)^{-\frac{1}{8}} h(x_*)^{-\frac{1}{4}}\int_{x_a}^\infty dx\frac{ \sqrt{G(x)}}{ \sqrt{e^{-\frac{1}{2}\phi(x)} M(x)^{-\frac{1}{4}} h(x)^{-\frac{1}{2}}-e^{\frac{1}{2}\phi(x_*)} M(x_*)^\frac{1}{4} h(x_*)^\frac{1}{2}}} \\
   &\approx \frac{\gamma e^{-\phi_{IR}/4}}{c L_{IR}} x_* \int_{x_a}^\infty dx e^{\frac{1}{4}\phi(x)} M(x)^\frac{1}{8} h(x)^\frac{1}{4}\sqrt{G(x)} \ .
\end{align}
Since the integral converges and is independent of $x_*$ we have found that $l_{UV} \sim x_*$ and that altogether $l\sim 1/x_*$ as $x_*\to 0$. The regularized geodesic length works out in a similar manner
\bea
   \mathcal{L}^{reg} & = & 2 \left(\int_{x_*}^{x_a}+\int_{x_a}^\infty\right) \Bigg(\frac{M(x)^{-\frac{1}{8}} e^{-\frac{\phi(x)}{4}} h(x)^{-\frac{1}{4}} \sqrt{G(x)}}{ \sqrt{1-e^{\frac{1}{2}(\phi(x)-\phi(x_*))} \left( \frac{M(x)}{M(x_*)} \right)^\frac{1}{4} \left( \frac{h(x)}{h(x_*)} \right)^\frac{1}{2}}} \nonumber \\
    & & \qquad\qquad\qquad\qquad - \frac{\chi(b)}{2} \frac{N^{1/4}}{\lambda^{1/16}} \left( \frac{r_q^2 \Theta}{\sqrt\lambda} \right)^{1/4} x^{-1+\frac{1}{2b}} \Bigg)dx - \frac{\chi(b)}{2} \frac{N^{1/4}}{\lambda^{1/16}} \left( \frac{r_q^2 \Theta}{\sqrt\lambda} \right)^{1/4} x_*^\frac{1}{2b} \nonumber \\
   & = & e^{-\frac{\phi_{IR}}{4}} L_{IR} \log \left( \frac{2 x_a}{x_*} \right) \nonumber\\
   & & + 2 \int_{x_a}^\infty dx \left( M(x)^{-\frac{1}{8}} e^{-\frac{\phi(x)}{4}} h(x)^{-\frac{1}{4}} \sqrt{G(x)}  - 
   \frac{\chi(b)}{2} \frac{N^{1/4}}{\lambda^{1/16}} \left( \frac{r_q^2 \Theta}{\sqrt\lambda} \right)^{1/4} x^{-1+\frac{1}{2b}} \right) \ .
\eea
The only difference to the commutative ABJM case \cite{Bea:2013jxa} is that the last integral is different. The effect of this is only that the normalization ${\mathcal N}'$ of the IR two-point function will be different. 
Otherwise the CFT result is recovered
\be
   \langle \mathcal{O}(x)\mathcal{O}(y) \rangle_{IR} = \frac{\mathcal{N}'}{(r_q l/\sqrt{\lambda})^{2\Delta_{IR}}} \ ,
\ee
where
\be
   \Delta_{IR} = m L_{IR} e^{-\phi_{IR}/4} \ .
\ee


\section{Holographic entanglement entropy}\label{sec:EE}
In this section we begin to study the holographic entanglement entropy in the non-commutative Chern-Simons ABJM gauge theory and, in particular, focus on the effects that the flavor degrees of freedom bring in. For the most parts we will follow \cite{Barbon:2008ut,Fischler:2013gsa,Karczmarek:2013xxa} and make comments where we have novel aspects. We will only focus on the case where there is a non-trivial dependence on only one non-commutative coordinate. This singles out strip geometries.

Let us consider the prescription of \cite{Ryu} to compute the entanglement entropy in a region A of the space of a quantum field theory using its gravity dual theory. The formula for the holographic entanglement entropy of region A is
\begin{equation}
   S_A=\frac{1}{4 G_N^{10}} \int_\Sigma dx^8 e^{-2\Phi} \sqrt{\det(\hat{g}_\Sigma)},
  \label{ncentanglemententropy}
\end{equation}
where $\hat{g}_\Sigma$ is the string frame metric induced on the codimension 2 minimal bulk surface whose boundary is the entangling surface of A and has the same holonomy of A. The bulk surface $\Sigma=\gamma_A\times M_6$ is factorized to a two-dimensional part $\gamma_A$ in the external dimensions and a compact manifold $M_6$ filling the internal space of the background.

We can rewrite expression (\ref{ncentanglemententropy}) in the Einstein frame and in terms of the four-dimensional Newton constant $G_N^{(4)}=G_N^{(10)}/  \int_{M_6} e^{-\frac{3}{2} \phi} dx^6 \sqrt{\det(\hat{g}_6)}$ as
\begin{equation}
  S_A=\frac{{\text{Area}}(A)}{4 G_N^{4}} \ ,
  \label{ncentanglemententropyA}
\end{equation}
where ${\text{Area}}(A)=\int_{\gamma_A} dx^2 \sqrt{h}$ is the volume of $\gamma_A$. We are going to compute this non-local observable in the non-commutative background we are studying. 

We choose as region A an infinite strip of coordinate width $l$ in the $(x_1,x_2)$-plane, $A=\{(x_1,x_2) \in \mathbb{R}^2, -\frac{l}{2} \leq  x_1 \leq \frac{l}{2} \}$. We want to determine the minimal surface area $\gamma_A$. As the problem is invariant under translations along the $x_2$-direction, the problem translates in determining the profile $x=x(x_1)$ that makes the surface minimal. Starting from the ten-dimensional metric (\ref{nc10dmetric}), the induced metric on $\gamma_A\times M_6$ is
\begin{equation}
   \hat{g}_\Sigma = h(x)^{-1/2}\left( \left( M(x) +G(x) x'^2 \right)dx^2_{1}+M(x)dx^2_{2}  \right)+h(x)^{1/2}\left[ e^{2f(x)} g_{\mathbb{S}^4}+e^{2g(x)}g_{\mathbb{S}^2} \right] \ ,
  \label{nc8dmetricinducedonA}
\end{equation}
where $G(x)=\frac{1}{x^2} h(x) e^{2g(x)}$ and $x'=\frac{\partial x}{\partial x_1 }$. The determinant of this metric is
\begin{equation}
   \det(\hat{g}_\Sigma)=h^2 M (M+G x'^2) e^{8f+4g} \det(g_{\mathbb{S}^4}) \det(g_{\mathbb{S}^2}) \ .
  \label{nc8ddeterminantmetricinducedonA}
\end{equation}
Expressing the dilaton in terms of the commutative one, and integrating over the internal manifold, (\ref{ncentanglemententropy}) becomes
\begin{equation}
  S_A=\frac{V_6 L_2}{4 G_{N}^{(10)}} \int_{-\frac{l}{2}}^{\frac{l}{2}} dx_1 \sqrt{H(x)} \sqrt{1+\frac{G(x)}{M(x)}  x'^2} \ ,
  \label{ncentanglemententropycomputations}
\end{equation}
where $H(x)= h^2 e^{8f+4g-4\phi}$, $L_2=\int dx_2$, and $V_6=32\pi^3/3$ is the volume of $\mathbb{CP}^3$ with the standard Fubini-Study metric. As the Lagrangian in (\ref{ncentanglemententropycomputations}) does not explicitly depend on $x_1$, we have a conserved quantity
\begin{equation}
  x' \frac{\partial L}{\partial x'} - L = \frac{H(x)}{\sqrt{1+\frac{G(x)}{M(x)} x'^2}} \ .
  \label{nceehamiltonian}
\end{equation}
The constant can be fixed by imposing the boundary condition $x'(x_*)=0$, that is, that the brane embedding is smooth at the tip $x=x_*$. This condition can be solved for $x'$ yielding
\begin{equation}
  x'=\pm \sqrt{\frac{M(x)}{G(x)}} \sqrt{\frac{H(x)}{H(x_*)}-1} \ .
  \label{nceehamiltonian1}
\end{equation}
Therefore, the width of the strip is given by
\begin{equation}
   l = \int_{-l/2}^{l/2}dx_1 = 2 \sqrt{H(x_*)} \int_{x_*}^{x_{max}} \frac{\sqrt{G(x)}}{\sqrt{M(x)} \sqrt{H(x)-H(x_*)}} dx \ .
  \label{nceehamiltonian2}
\end{equation}
Physical distances on the field theory side are measured with the open string metric (\ref{eq:open_string_metric}) which implies that, after stripping off the conformal $AdS$-factor, the boundary is governed by the usual Minkowski metric. Here the distinction between closed and open string metrics is important because now we can see that even though the bulk metric is not asymptotically $AdS$, equation (\ref{nceehamiltonian2}) still corresponds to the physical width of the entangling region. We will regularize the entanglement entropy by introducing a cutoff at $x=x_{max}$ and we choose to introduce the same cutoff in (\ref{nceehamiltonian2}).

\begin{figure}
   \centering
   \includegraphics[width=1.0\textwidth]{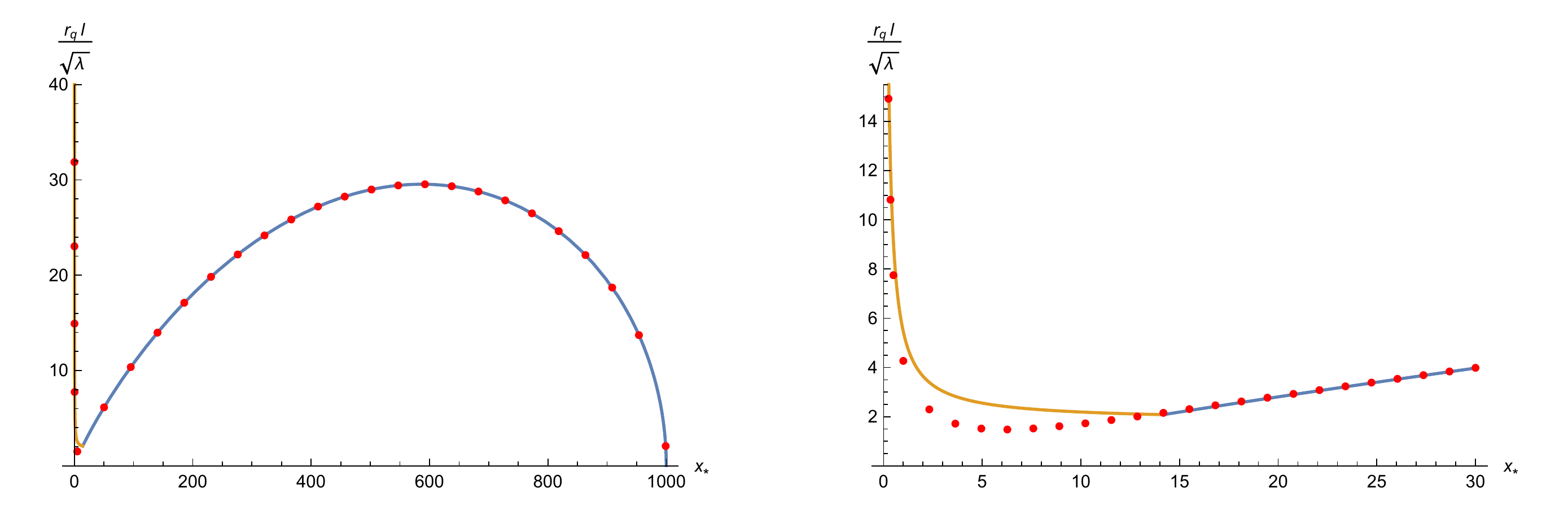}
   \caption{Strip width as a function of tip position $x_*$ for $r_q^2\Theta /\sqrt\lambda=0.1$ and $\hat{\epsilon}=1$. For large $r_q l/\sqrt\lambda$ strips have a small $x_*$ and for small $r_q l/\sqrt\lambda$ they have a large $x_*$ with a phase transition in between in the region where there are three possible embeddings for a given value of $r_q l/\sqrt\lambda$. Red dots represent numerical data and solid curves correspond to analytical IR/UV-expansions.}
   \label{fig:strip_l_vs_xs}
\end{figure}

Fig.~\ref{fig:strip_l_vs_xs} shows the behavior of $l(x_*)$ showing that for very large $l$ the strips probe the deep IR and for very small $l$ the deep UV. In the middle there is a range in $l$ that has three different candidate surfaces for the absolute minimal surface. There is a phase transition at some $l=l_{crit}$ where minimal area strips jump from those probing the IR to those probing the UV. We wish to compute the point $l_{crit}$, where the phase transition happens. For this we need to compute the entanglement entropy. We plug in the embedding (\ref{nceehamiltonian1}) into (\ref{ncentanglemententropycomputations}) to obtain
\begin{align}
   S_A = \frac{V_6 L_2}{2 G_N^{(10)}} \int_{x_*}^{x_{max}} dx \frac{\sqrt{G(x)} H(x)}{\sqrt{M(x)}\sqrt{H(x)-H(x_*)}} \ . \label{eq:strip_S_integral}
\end{align}
The divergences coming from correlations across $\partial A$ are dealt with by the introduction of a cutoff. We define the regularized entanglement entropy for numerical convenience
\begin{align}
   S_A^r = S_A - S_A^{div} \ , \label{eq:strip_S_regularized}
\end{align}
where we have subtracted the cutoff divergence
\begin{gather}
   \frac{\lambda S_A^{div}}{r_q N^2 L_2} = \frac{\xi(b)\kappa^3}{9\sqrt{2}\pi^2\sigma(b)^2} \frac{r_q^2 \Theta}{\sqrt{\lambda}} x_{max}^{3/b} \left( 1+ \frac{3 C_2}{3-2b} x_{max}^{-2} \right) \ , \label{eq:strip_S_div}
\end{gather}
where 
\begin{align}
   C_2 = \frac{3 (b-1) (3 b-5) (b (4 b (4 b-5)-25)+30)}{2 b^2 (2 b+3) ((b-13) b+15)}
\end{align}
is a flavor dependent constant, $\sigma(b)$ is the screening function (\ref{eq:screening_func}), and
\begin{align}
   \xi (b) = \frac{1}{16} \frac{q_0^{5/2}(1+\hat{\epsilon}+q_0)^4}{(2-q_0)^{1/2}(q_0+(1+\hat{\epsilon}) q_0-(1+\hat{\epsilon}))^{7/2}} \ ,
\end{align}
where $q_0$ is a $b$-dependent constant given by
\begin{align}
   q_0 = \frac{b}{2-b} \ .
\end{align}

\begin{figure}
   \centering
   \includegraphics[width=0.7\textwidth]{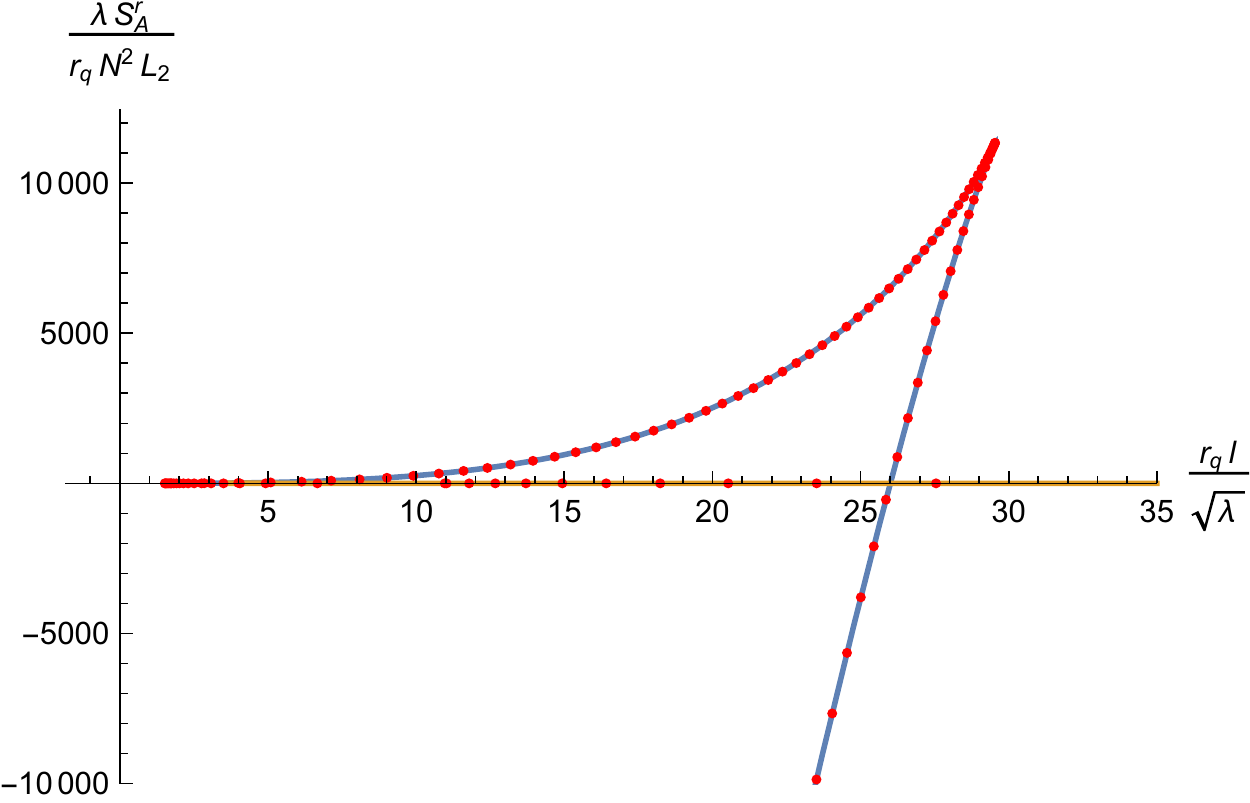}
   \caption{Regularized entanglement entropy for parameter values $\Theta=0.1$, $\hat{\epsilon}=1$, and $x_{max}=1000$, computed numerically from (\ref{eq:strip_S_integral}) and (\ref{eq:strip_S_regularized}). There is a phase transition between the IR-branch (the one following the horizontal axis) and the UV-branch when $l=l_{crit}$.}
   \label{fig:strip_S_vs_l}
\end{figure}

\begin{figure}
   \centering
   \includegraphics[width=1.0\textwidth]{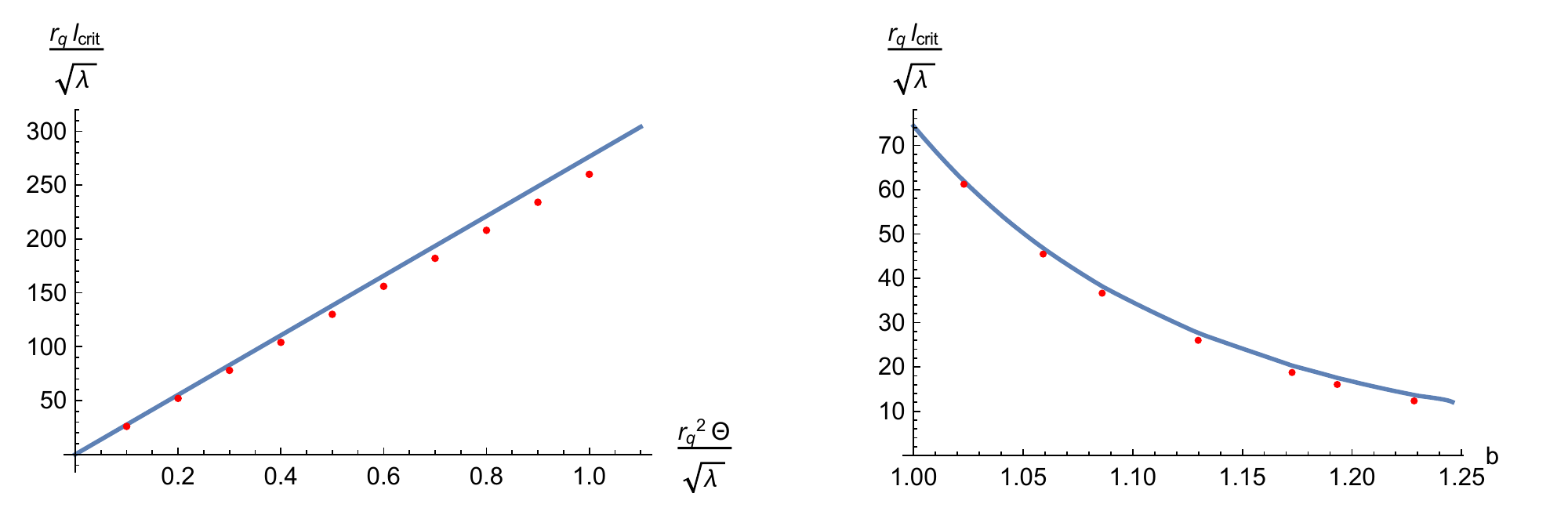}
   \caption{$l_{crit}$ as a function of the non-commutativity parameter $\Theta$ and flavor parameter $b$. Red dots correspond to numerical data and solid curves represent analytical formulas. In the left figure we have fixed $b=1$ and in the right figure we have fixed $\frac{r_q^2 \Theta}{\sqrt{\lambda}}=0.1$. For fixed $b$, we find a linear dependence on $\Theta$. For fixed $\Theta$, we see a more complicated dependence on $b$. We have set $x_{max}=1000$.}
   \label{fig:lcrit}
\end{figure}

We evaluate the integrals (\ref{nceehamiltonian2}) and (\ref{eq:strip_S_integral}) numerically to investigate the phase transition. Results are shown in Fig.~\ref{fig:strip_S_vs_l}. We wish to find how the phase transition, represented by $l_{crit}$, depends on $\Theta$ and $\hat{\epsilon}$. Results are presented in Fig.~\ref{fig:lcrit}. $\Theta$ and $b$ have opposite effects on $l_{crit}$. As in \cite{Karczmarek:2013xxa}, increasing $\Theta$ causes the phase transition to happen for wider strips. Interestingly, increasing the number of flavors makes the phase transition to occur at smaller strip widths. The flavor degrees of freedom in a sense make the field theory more local towards the UV and we can probe smaller distances. However, even in the limit of infinite flavors, the minimum distance never vanishes and one always lands on the non-local regime in the deep UV.

We now turn our attention to gaining some analytical intuition over the function $l_{crit}(b,\Theta)$. In order to accomplish this, the crucial observation is that the phase transition seems to happen between strips that probe the fairly deep IR ($x_*<1$) and strips that probe the asymptotic region ($x_* \gg 1$) where background running is no longer relevant. This suggests that $l_{crit}(b,\Theta)$ could be computed by comparing IR and UV expansions for the strip entanglement entropy, where we have some analytical control.

The IR limit can be computed by following \cite{Bea:2013jxa} with the difference that we are employing an UV cutoff $x=x_{max}$. IR behavior is the same as in the commutative case because $M(x\to 0)=1$. In the end we obtain first for $l(x_*,x_{max})$
\begin{align}
   \frac{r_q l(x_*,x_{max})}{\sqrt{\lambda}} = \frac{8\pi^{5/2}}{\Gamma\left( \frac{1}{4} \right)^2(1+\hat{\gamma})} \frac{1}{x_*} + \mathcal{O}(x_*^2) \ .
\end{align}
This expression is plotted in Fig.~\ref{fig:strip_l_vs_xs} as a orange solid curve. Finally we have for the entanglement entropy
\begin{align}
   S_A^{IR} (l) = - \frac{4\sqrt{2}\pi^3}{3\Gamma\left( \frac{1}{4} \right)^4} \frac{N^2 r_q L_2}{\lambda} \frac{\sqrt{\lambda}}{r_q l} + S_A^\infty (x_{max}) + S_A^{div}(x_{max}) \ , \label{eq:strip_S_IR}
\end{align}
where $S_A^{div}(x_{max})$ is given in (\ref{eq:strip_S_div}) and
\begin{align}
   S_A^\infty (x_{max}) = \frac{V_6 L_2}{2 G_N^{(10)}} \int_0^{x_{max}} \left( \sqrt{ \frac{G(x)H(x)}{M(x)}} - \sqrt{ \frac{G_\infty H_\infty}{M_\infty}} x^{-1+3/b}(1+C_2 x^{-2}) \right) dx
\end{align}
which is convergent when $x_{max}\to\infty$. We recover the usual $S\sim l^{-1}$ behavior of $AdS_4$, as expected.

As our spacetime is not asymptotically $AdS_4$, we expect to see something more interesting in the UV calculation. We start by using the UV expansions for metric functions to obtain first $l(x_*,x_{max})$ in the limit where both the tip coordinate $x_*$ and cutoff $x_{max}$ are assumed to be large,
\begin{align}
   \frac{r_q l(x_*,x_{max})}{\sqrt{\lambda}} = 2 \frac{r_q^2 \Theta}{\sqrt{\lambda}} x_*^{1/b} \left( \frac{\pi^{3/2}}{2\sqrt{2}\Gamma\left( \frac{3}{4}\right)^2} - \left( \frac{x_*}{x_{max}} \right)^{1/b} {}_2 F_1 \left( \frac{1}{4}, \frac{1}{2}; \frac{5}{4}; \left( \frac{x_*}{x_{max}} \right)^{4/b} \right)  \right) \ . \label{eq:strip_l_UV}
\end{align}
This expression is the blue solid curve plotted on top of numerical data in Fig.~\ref{fig:strip_l_vs_xs}. Again by using UV expansions for background functions we can integrate to find $S_A^{UV}(x_*,x_{max})$,
\begin{align}
   \frac{ \lambda S_A^{UV,r}(x_*,x_{max})}{r_q N^2  L_2} & = \frac{\xi(b) \kappa^3}{9\sqrt{2}\pi^2\sigma(b)^2} \frac{r_q^2 \Theta}{\sqrt{\lambda}} x_*^{3/b} \left[ \left( \frac{x_*}{x_{max}} \right)^{-4/b}\sqrt{ \left( \frac{x_*}{x_{max}} \right)^{-4/b}-1 } - \left( \frac{x_*}{x_{max}} \right)^{-3/b} \right. \nonumber \\
   & - \left. \text{Im} F \left( \arcsin \left( \left(\frac{x_*}{x_{max}}\right)^{-1/b} \right)  \Bigg| -1 \right)  \right] \ . \label{eq:strip_S_UV}
\end{align}

The phase transition point is then found by solving $S_A^{UV}(l_{crit})=S_A^{IR}(l_{crit})$. We can further approximate $S_A^{UV}(l_{crit})\approx 0$ because in (\ref{eq:strip_S_UV}) $x_*,x_{max}\gg 1$ implying $S_A^{UV}(l_{crit}) \gg S_A^{IR}(l_{crit})$. First we solve $S_A^{UV}(x_*,x_{max})=0$ for the point $x_*$ corresponding to the phase transition and then obtain $l_{crit}$ from formula (\ref{eq:strip_l_UV}). Note that the equation $S_A^{UV}(x_*,x_{max})=0$ does not depend on $\Theta$. This means that $l_{crit}\sim \Theta$ as can be noted by (\ref{eq:strip_l_UV}). In Fig.~\ref{fig:lcrit} we will show this analytic result for $l_{crit}(b,\Theta)$ and compare it with numerical results finding nice agreement.


\section{Conclusions}\label{sec:conclusions}

The unquenched massive ABJM provides a unique laboratory to investigate how the Chern-Simons scalar matter together with fundamental degrees of freedom behave under the renormalization scale transformations between distinct conformal fixed points at the UV and at the IR. In this paper, we refined this construction presented in \cite{Bea:2013jxa} and introduced a parameter $\Theta$ in the field theory side rendering the commutator of the coordinate operators non-vanishing. We presented the gravity dual of this system and discussed at length the novel aspects that the dynamical quarks had on the geometry in comparison to other frameworks in the literature. Having discussed the background geometry, we then turned into studying several select observables.

Particular examples that we studied consisted of dual observables to spinning and/or hanging strings and we obtained results in parallel with the expectations from the other non-commutative backgrounds. The emphasis in our analysis was on the flavor effects from the dynamical quarks in the background. We then further considered the holographic entanglement entropy of strip geometries, and, again, first found close resemblance to other investigations with no flavors and then invested words on explaining the flavor effects. One important take-home message of our studies in both of these cases is that the extra flavor degrees of freedom in the system will alter the minimum fundamental distance scale beyond which one cannot probe the system at hand.

While all these results are interesting on their own right and could also find applications elsewhere, we wish to emphasize that our main motivation for this work was to establish the gravity dual of non-commutative unquenched massive flavored ABJM, which we can use as a platform for more ambitious questions. In particular, as discussed in the introduction of this paper, we are very interested in understanding the fuzzy granular nature of the electrons which are subject to strong magnetic fields. In our previous work in this field \cite{Bea:2014yda} we constructed the holographic dual of a quantum Hall state in this particular field theory. To supply answers to the underlying problems related to the non-commutative behavior of electrons in Chern-Simons theories, we are planning to continue our efforts and combine these two works in the future.

Other generalizations and further outgrowths consist of incorporating temperature effects and further study the embedded probe D6-branes with external magnetic gauge fields \cite{Jokela:2013qya} to make contact with the proposed ``inverse'' magnetic catalysis to see how the non-commutative parameter changes the picture. While in all these approaches the gauge parameters (for example the non-commutative angle $\Theta$) are to be taken fixed, we would like to ask if something qualitatively different behavior is to be expected if we allow them being dynamical \cite{Witten:2003ya, BurgessandDolan,Jokela:2013hta,Brattan:2013wya,Ihl:2016sop}? In fact, the lowest excitation of the quantum Hall state should really be, depending on the form of the fluctuation, either magneto-roton or anyonic in nature. This expectation has been met in other cousin holographic quantum Hall proxies \cite{Bergman:2010gm,Jokela:2010nu,Jokela:2011eb,Jokela:2011sw,Jokela:2013hta,Jokela:2014wsa} and we hope to establish this also in the setup proposed here.

\medskip
\paragraph{Acknowledgments}
We would like to thank David Mateos, Ioannis Papadimitriou, and Jorge Russo for useful discussions. Y.~B. is supported by grants MEC FPA2013-46570-C2-1-P, MEC FPA2013-46570-C2-2-P, MDM-2014-0369 of ICCUB, 2014-SGR-104, 2014-SGR-1474, CPAN CSD2007-00042 Consolider-Ingenio 2010, and ERC Starting Grant HoloLHC-306605. N.J.~and A.P.~have been supported in part by the Academy of Finland grant no.~1303622. A.P.~also acknowledges support from the Magnus Ehrnrooth foundation.  A.~V.~R. is funded by the Spanish grant FPA2014-52218-P, by Xunta de Galicia (GRC2013-024), and by FEDER.


\appendix
\vskip 1cm

\setcounter{equation}{0}
\medskip

\section{Details of the background}\label{sec:DetailsBackground} 

In this appendix we specify the coordinate system that we use to write down the metric and the forms of the background. Let $\omega^i$ ($i=1,2,3$) be the $SU(2)$ left-invariant one-forms, which satisfy $\omega^i=\frac{1}{2} \epsilon_{ijk} \omega^j \wedge \omega^k$. Using these one forms and a new coordinate $\alpha$, the metric on the four sphere $\mathbb{S}^4$ can be written as
\begin{equation}
ds_{\mathbb{S}^4}^2=d\alpha^2+\frac{\sin^2 \alpha}{4} \left[ (\omega^1)^2+(\omega^2)^2+(\omega^3)^2 \right]~~,
\label{4sphere_coordinates}
\end{equation}
where $0 \leq \alpha<\pi$. Now, we can parameterize the coordinates of the fibered $\mathbb{S}^2$-sphere by
\begin{equation}
z^1=\sin \theta \cos \varphi~~, ~~~~~~~z^2=\sin \theta \sin \varphi~~, ~~~~~~~z^3= \cos \theta~~, 
\label{fibered_S2_sphere_coordinates}
\end{equation}
where $0 \leq  \theta<\pi$, $0 \leq  \varphi<2 \pi$.  Then, the fibered part of the metric can be written as
\begin{equation}
\left( dz^i-\sin(\frac{\alpha}{2}) \epsilon_{ijk} \omega^j z^k \right)^2=(E^1)^2+(E^2)^2~~, ~~~~ i=1,2,3~~, 
\label{fibered_S2_sphere_one_formsE1E2}
\end{equation}
where $E^1$ and $E^2$ are the following one-forms
\begin{align}
E^1 &=d \theta+\sin^2 (\frac{\alpha}{2})\left[ \sin \varphi \ \omega^1-\cos \varphi \ \omega^2 \right] \\
E^2 &= \sin  \theta \left[   d \phi -\sin^2 (\frac{\alpha}{2}) \omega^3   \right] + \sin^2 (\frac{\alpha}{2}) \cos \theta \left[ \cos \varphi \omega^1+\sin \varphi \omega^2 \right] \ .
\label{E1E1one_forms}
\end{align}
Furthermore, to write down the forms of the background, we define
\begin{align}
{\cal S}^{\alpha} &= d\alpha \\
{\cal S}^1 &=\frac{\sin\alpha}{2}\left( \sin \varphi \ \omega^1-\cos \varphi \ \omega^2 \right) \\
{\cal S}^2 &= \frac{\sin\alpha}{2}\left[ \sin \theta \ \omega^3- \cos \theta \left( \cos \varphi \ \omega^1  + \sin \varphi \ \omega^2 \right) \right] \\
{\cal S}^3 &=  \frac{\sin\alpha}{2}\left[ -\cos \theta \ \omega^3- \sin \theta \left( \cos \varphi \ \omega^1  + \sin \varphi \ \omega^2 \right) \right] \ ,
\label{S1S2S3salpha_one_forms}
\end{align}
and in terms of these one-forms, the metric on the four-sphere reads $ds_{\mathbb{S}^4}^2=({\cal S}^{\alpha})^2+\sum _i^3 ({\cal S}^i)^2$.

In order to define the specific embedding for the D6-brane in Appendix~\ref{sec:kappa}, we write down the SU(2)-left invariant one-forms in the following coordinates
\begin{align}
\omega^1 &=  \cos (2 \beta) d \hat{\theta} + \sin  (2 \beta) \sin \hat{\theta} d\hat{\varphi}  \\
\omega^2 &=  \sin (2 \beta) d \hat{\theta} - \cos  (2 \beta) \sin \hat{\theta} d\hat{\varphi}  \\
\omega^3 &= 2d\beta +\cos \hat \theta  d\hat{\varphi}   \ ,
\label{SU2_invariant_one_forms}
\end{align}
where $0\leq\beta,\hat{\theta}\leq\pi$, $0\leq\hat{\psi}\leq4\pi$.

\section{Supersymmetry of the background}\label{sec:SUSY2} 

In this appendix we prove that the non-commutative unquenched massive flavored ABJM solution presented in Section~\ref{sec:TsT} is ${\cal N}=1$ supersymmetric. We check that the fermionic sector of the theory remains vanishing under supersymmetric transformations. From reference \cite{dilatinogravitino}, eq. (80), (81), and (82), the dilatino and gravitino variations in the string frame for Type IIA supergravity are
\begin{align}
\delta_{\epsilon} \lambda&=\left[\frac{1}{2} \left( \Gamma^a \partial_a \Phi + \frac{1}{2 \cdot 3!} H_{mnp} \Gamma^{mnp} \Gamma_{11} \right) +\frac{e^{\Phi}}{8} \left( \frac{3}{2!} F_{mn} \Gamma^{mn} \Gamma_{11} -\frac{1}{4!} F_{mnpq} \Gamma^{mnpq} \right) \right] \epsilon  ~~~~~~~\label{dilatinoeq} \\
\delta_{\epsilon} \Psi_{\mu}&=\left[\bigtriangledown_{\mu} +\frac{1}{8}H_{\mu ab} \Gamma^{ab} \Gamma_{11} -\frac{1}{8} e^{\Phi} \left( \frac{1}{2} F_{ab} \Gamma^{ab} \Gamma_{11} + \frac{1}{24} F_{abcd} \Gamma^{abcd} \right) \Gamma_{\mu} \right] \epsilon \ .  \label{gravitinoeq}
\end{align}
A remarkable property of the TsT rotated theory is that the BPS equations are the same as in the unrotated case. Then, solutions in the commutative case are solutions of the non-commutative case. In particular, the unquenched massive flavored ABJM solution is rotated into its non-commutative version. From \cite{Bea:2013jxa} , eq. (3.6), (A.1), (A.2), and (A.3), the explicit expression of the BPS equations are\footnote{By $\phi(x)$ we mean the dilaton of the commutative theory, not to be confused with $\Phi(x)$, the dilaton of the non-commutative theory. The precise relation between them is (\ref{ncdil}). The functions $\eta$, $h$, $g$, $f$, and $K$ are the same in both theories. Notice that $e^{2f}=q e^{2g}$, where $q$ we denote by the squashing factor.}
\begin{align}
\frac{x}{e^g}\phi'&=- \frac{3k}{8} e^{\phi} h^{-\frac{1}{4}} (e^{-2g}-2 \eta e^{-2f})-\frac{e^{\phi}}{4} K h^{\frac{3}{4}} \label{BPS001}\\
\frac{x}{e^g}h'&= \frac{k}{2} e^{\phi} h^{\frac{3}{4}} (e^{-2g}-2 \eta e^{-2f})-e^{\phi} K h^{\frac{7}{4}} \label{BPS002} \\
\frac{x}{e^g}f'&=\frac{k}{4} e^{\phi} h^{-\frac{1}{4}} ( \eta e^{-2f}-e^{-2g})+e^{-2f+g} \label{BPS003}\\
\frac{x}{e^g}g'&=\frac{k}{2} e^{\phi} h^{-\frac{1}{4}} \eta e^{-2f}+e^{-g}-e^{-2f+g} \ . \label{BPS004}
\end{align}
Let us choose the following basis of frame one-forms for the metric (\ref{nc10dmetric})
\begin{align}
e^0&=h^{-\frac{1}{4}} dx_0 \ & e^1&= \sqrt{M}  h^{-\frac{1}{4}} dx_1 \  & e^2&= \sqrt{M}  h^{-\frac{1}{4}} dx_2 \  &  e^3&=  h^{\frac{1}{4}} \frac{e^g}{x} dx \ & e^4 &= h^{\frac{1}{4}} e^f {\cal S}^{\xi} ~~~~ \nonumber \\  e^5 &= h^{\frac{1}{4}} e^f {\cal S}^{1} & e^6 &= h^{\frac{1}{4}} e^f {\cal S}^{2} &  e^7&= h^{\frac{1}{4}} e^f {\cal S}^{3} &
e^8 &= h^{\frac{1}{4}} e^f E^{1} & e^9 &= h^{\frac{1}{4}} e^f E^{2}  \ .
\label{frame_one_forms}
\end{align}
Let us impose that the dilatino variation (\ref{dilatinoeq}) is vanishing. Using the K\"ahler projections
\begin{equation}
\Gamma_{47} \epsilon = \Gamma_{89} \epsilon=\Gamma_{56} \epsilon~,
\label{kahlerproj}
\end{equation}
and expressing the dilaton in terms of the commutative one we arrive at
\begin{equation}
\delta_{\epsilon} \lambda= \Gamma_3 \left[ \Lambda_1 + \Lambda_2 \Gamma_{12} \Gamma_{11} + \Lambda_3 \Gamma_{012} + \Lambda_4 \Gamma_{0} \Gamma_{11}  \right] =0~,
\label{dilatinovarequaltozero}
\end{equation}
where
\begin{align}
\Lambda_1&= \frac{x}{e^g} \left(\frac{1}{2}h^{-\frac{1}{4}} \phi '  + \frac{1}{4} h^{-\frac{9}{4}} \Theta^2 h' M \right) \\
\Lambda_2&=\frac{\Theta M x h'}{4 e^g h^{\frac{7}{4}}} \\
\Lambda_3&=-\frac{e^{\phi} \sqrt{M} h^{\frac{1}{2}}}{8} \left( \frac{3}{2} k h^{-1} (e^{-2g}-2 \eta e^{-2f})+ K\right) \\
\Lambda_4&=-\frac{\Theta e^{\phi} \sqrt{M} }{8} \left( \frac{k}{2} h^{-1}(e^{-2g}-2 \eta e^{-2f}) +3 K  \right) \ .
\label{dilatinocomp}
\end{align}
From (\ref{dilatinovarequaltozero}) we obtain
\begin{equation}
\Gamma_{012} \epsilon= \left( - \frac{\Lambda_1 \Lambda_3 +\Lambda_2 \Lambda_4}{\Lambda_3^2+\Lambda_4^2} + \frac{\Lambda_1 \Lambda_4 - \Lambda_2 \Lambda_3}{\Lambda_3^2+\Lambda_4^2} \Gamma_{12} \Gamma_{11}\right) \epsilon~.
\label{dilatinocomp2}
\end{equation}
Using equations (\ref{BPS001}) and (\ref{BPS002}) we arrive at
\begin{equation}
\Gamma_{012}  \epsilon =\sqrt{M} \left( -1+ \frac{\Theta}{\sqrt{h}} \Gamma_{12} \Gamma_{11} \right) \epsilon \ .
\label{rotatedprojection0}
\end{equation}
Defining
\begin{equation}
\beta_{\Theta} =\arctan( \frac{\Theta}{\sqrt{h}}) 
\label{definition_of_beta_theta}
\end{equation}
the equation (\ref{rotatedprojection0}) can be rewritten as
\begin{equation}
\Gamma_{012}  \epsilon=\left( -\cos(\beta_{\Theta} )+ \sin(\beta_{\Theta} ) \Gamma_{12} \Gamma_{11} \right) \epsilon= - e^{-\beta_{\Theta} \Gamma_{12} \Gamma_{11} } \epsilon \ .
\label{rotatedprojection1}
\end{equation}
A solution to this equation is
\begin{equation}
 \epsilon=  e^{\frac{\beta_{\Theta}}{2} \Gamma_{12} \Gamma_{11} } \epsilon_0~,
\label{solutiontorotatedprojection1}
\end{equation}
where $\epsilon_0$ satisfies
\begin{equation}
\Gamma_{012}  \epsilon_0= - \epsilon_0 \ .
\label{epsilon0}
\end{equation}
Let us impose that the gravitino variation (\ref{gravitinoeq}) is vanishing. For $\mu=0,1,2$ the equations are satisfied when K\"ahler projections (\ref{kahlerproj}), projection (\ref{rotatedprojection0}), and equation (\ref{BPS002}) are used. For $\mu=x$ using K\"ahler projections (\ref{kahlerproj}) and projection (\ref{rotatedprojection0}) we arrive at
\begin{equation}
h^{-\frac{1}{4}}  \frac{x}{e^g} \partial_x \epsilon +\left[ \frac{1}{4} \frac{\Theta M x h'}{e^g h^{\frac{7}{4}}} \Gamma_{12} \Gamma_{11} - \frac{e^{\phi} M}{8} (h+ \Theta^2) \left(  \frac{K}{\sqrt{h}} -\frac{k}{2} h^{-\frac{3}{2}}  (e^{-2g}-2 \eta e^{-2f}) \right)  \right] \epsilon=0~.
\label{Killingspinoreq}
\end{equation}
Using (\ref{solutiontorotatedprojection1}),
\begin{equation}
 \frac{x}{e^g} \partial_x \epsilon_0 =\frac{e^{\phi} h^{-1}}{8} \left(  h^{\frac{7}{4}} K -\frac{k}{2} h^{\frac{3}{4}}  (e^{-2g}-2 \eta e^{-2f}) \right) \epsilon_0=- \frac{x}{e^g} \frac{h^{-1}}{8} h' \epsilon_0~,
\label{Killingspinoreq1}
\end{equation}
and integrating the last expression we obtain
\begin{equation}
\epsilon_0 =h^{-\frac{1}{8}} \eta_0~,
\label{Killingspinoreq2}
\end{equation}
where $\eta_0$ is a $x$-independent spinor. For $\mu=4,5,6,7$, the equations are satisfied if we impose K\"ahler projections (\ref{kahlerproj}), projection (\ref{rotatedprojection0}), equations (\ref{BPS002}) and (\ref{BPS003}), and a new projection
\begin{equation}
\Gamma_{3458} \epsilon=- \epsilon \ .
\label{thirdprojection}
\end{equation}
For $\mu=8,9$, the equations are satisfied if we impose K\"ahler projections (\ref{kahlerproj}), projections (\ref{rotatedprojection0}) and (\ref{thirdprojection}), and equations (\ref{BPS002}) and (\ref{BPS004}).

In conclusion, we impose the projections (\ref{kahlerproj}), (\ref{rotatedprojection0}), (\ref{thirdprojection}) on the Killing spinor and the BPS equations (\ref{BPS001})-(\ref{BPS004}) for the dilatino and gravitino variations to vanish, thus we conclude that the theory is ${\cal N}=1$ supersymmetric.


\section{Kappa symmetry analysis}\label{sec:kappa}

In this appendix we consider the kappa symmetry of a probe flavor D6-brane in our background. We consider a massive embedding with a generic mass independent of the mass of the background flavors. In addition, we will consider internal flux on the compact part of the worldvolume, as in \cite{Bea:2014yda}. This internal flux breaks parity, and allows to turn on a Wess-Zumino term crucial for the construction of the holographic quantum Hall system. The inclusion of this flux in the non-commutative set up will be useful in future work for the construction of a non-commutative quantum Hall system.

The kappa symmetry matrix for a D$p$-brane in the Type IIA theory is given by
\begin{equation}
d^{p+1} \zeta  \  \Gamma_{\kappa}=\frac{1}{\sqrt{-\det(\hat{g}+F)}}  e^F \wedge X \ , \label{kappasymmetricmatrixgeneral}
\end{equation}
where $F$ is defined as $F:={\cal F}-\hat{B_2}$ , $\hat{g}$ and $\hat{B_2}$ are the induced metric and NS two-form on the worldvolume, ${\cal F}$ is any additional gauge field living on the worldvolume of the brane, and $\zeta^{\alpha}$ ($\alpha=0,...,p$) are the worldvolume coordinates of the brane. The polyform matrix $X$ is
\begin{equation}
X=\sum_n  \gamma_{(2n+1)} \left( \Gamma_{11} \right)^{n+1} \ ,\label{polyformmatrix}
\end{equation}
where $\gamma_{(2n+1)}$ is the $(2n+1)$-form whose components are the antisymmetrized products of the induced Dirac matrices $\gamma_{\mu}$
\begin{equation}
 \gamma_{(2n+1)}=\frac{1}{(2n+1)!} \gamma_{\mu_1 ...\mu_{2n+1}} d\zeta^{\mu_1} \wedge ...\wedge d\zeta^{\mu_{2n+1}}  \ . \label{gammamatricesform}
\end{equation}
With these conventions, supersymmetric embeddings are those which satisfy $\Gamma_{\kappa} \epsilon=-\epsilon$, where $\epsilon$ is a Killing spinor of the background.

We consider the following embedding for the flavor D6-brane in our background. We define $\psi=\varphi-\hat{\psi}/2$, and choose for the worldvolume coordinates
\begin{equation}
 \zeta^{\alpha} =(x_0,x_1,x_2,x,\alpha,\beta,\psi) \ , \label{worldvolumecoordinates}
\end{equation}
and an embedding given by
\begin{equation}
\hat{\varphi} = \text{constant}~,  ~~~~~~~~~~   \hat{\theta} = \text{constant}~,  ~~~~~~~~~~~~~ \theta = \theta(x)  \ , \label{embeddingsusymassivebrane}
\end{equation}
with a non-trivial gauge field on the worldvolume of the probe D6-brane. The induced $B_2$ field is
\begin{equation}
\hat{B}_2=  \frac{\Theta M}{h} dx_1 \wedge dx_2 ~.
\label{embeddingsusymassivebrane2}
\end{equation}
Moreover, we also turn on an internal flux on the compact part of the worldvolume of the brane:
$${\cal A} =a(x)(d\psi + \cos \alpha d\beta)~,$$
\begin{equation}
{\cal F}= d{\cal A} =a'(x)dx\wedge(d\psi + \cos \alpha d\beta)-a(x) \sin \alpha \ d  \alpha \wedge d\beta   \ ,
\label{embeddingsusymassivebrane3}
\end{equation}
where $a(x)$ is an embedding function to be determined by the kappa symmetry condition. So, the total gauge field on the brane is
\begin{equation}
F =\frac{\Theta M}{h} dx_1 \wedge dx_2+a'(x)dx\wedge(d\psi + \cos \alpha d\beta)-a(x) \sin \alpha \ d  \alpha \wedge d\beta ~.
\label{embeddingsusymassivebrane4}
\end{equation}
We furthermore define
\begin{equation}
F_{x_1x_2}=\frac{\Theta M}{h}~, ~~~~~~~~ F_{x \psi}= a'  ~, ~~~~~~~~ F_{x \beta}= a' \cos \alpha ~, ~~~~~~~~ F_{\alpha \beta}= -a \sin \alpha ~.
\label{gauge_field_brane_components_definitions}
\end{equation}
The induced metric on the worldvolume of the flavor D6-brane is
\bea
 ds^2_7 & =& h^{-\frac{1}{2}}\left( -dx_0^2+M (dx_1^2+dx_2^2)\right)+h^{\frac{1}{2}}\left( \frac{1}{x^2}+\theta'(x)^2 \right) e^{2g}dx^2 \nonumber\\
 & &  +h^{\frac{1}{2}}\left( e^{2f} (d\alpha^2+\sin^2 \alpha d\beta^2)+e^{2g} \sin^2 (\theta(x)) (d\psi +\cos \alpha d \beta)^2 \right)  \ . \label{ncinducedmetricthetax}
\eea
The induced one-forms on the worldvolume are
\begin{align}
\hat{e}^0&= h^{-\frac{1}{4}} dx_0               &
\hat{e}^1&= \sqrt{M}  h^{-\frac{1}{4}} dx_1      &
\hat{e}^2&= \sqrt{M}  h^{-\frac{1}{4}} dx_2      \nonumber  \\ 
\hat{e}^3&= h^{\frac{1}{4}} \frac{e^g}{x} dx     &
\hat{e}^4&= h^{\frac{1}{4}} e^f d\alpha          &
\hat{e}^5&= 0                                     \nonumber \\
\hat{e}^6&= h^{\frac{1}{4}} e^f \sin  \alpha  \sin \theta d\beta     &
\hat{e}^7&= -h^{\frac{1}{4}} e^f \sin \alpha \cos \theta d\beta      &
\hat{e}^8&= h^{\frac{1}{4}} e^g \theta' dx                          \nonumber   \\
\hat{e}^9&= h^{\frac{1}{4}} e^g \sin \theta \left( d \psi + \cos\alpha d \beta \right)~,  & 
\label{induced_one_forms_worldvolume}
\end{align}
and the induced gamma matrices are:
$$\gamma_{x_0}=-h^{-\frac{1}{4}} \Gamma_0 ~~~~~~ \gamma_{x_1}= \sqrt{M}  h^{-\frac{1}{4}} \Gamma_1  ~~~~~~~ \gamma_{x_2}= \sqrt{M}  h^{-\frac{1}{4}} \Gamma_2 ~~~~~~ \gamma_{x} =  h^{\frac{1}{4}} \frac{e^g}{x} \left( \frac{1}{x} \Gamma_3+\theta' \Gamma_8 \right) $$
\begin{equation}
\gamma_{\alpha} = h^{\frac{1}{4}} e^f \Gamma_4~~~~~ \gamma_{\beta} = h^{\frac{1}{4}} e^f \sin  \alpha  \sin \theta \left( \Gamma_6 -\frac{\cos \theta}{\sin \theta} \Gamma_7 + \frac{\sin \alpha}{\cos \alpha}  e^ {g-f} \Gamma_9 \right)  ~~~~ \gamma_{\psi} = h^{\frac{1}{4}} e^g  \sin \theta \Gamma_9~.
\label{inducedgammamatrices1}
\end{equation}
We will determine the functions $\theta(x)$ and $a(x)$ that make the embedding supersymmetric. 

In our particular case, expression (\ref{kappasymmetricmatrixgeneral}) reads
\begin{equation}
d^7 \zeta \Gamma_{\kappa}=\frac{1}{\sqrt{-\det(\hat{g}+F)}} \left( \gamma_{(7)} + F \wedge \gamma_{(5)} \Gamma_{11} +\frac{1}{2} F \wedge F \wedge \gamma_{(3)}+\frac{1}{6} F \wedge F \wedge F \wedge \gamma_{(1)} \Gamma_{11} \right) ~~.
\label{kappasymmetricmatrix}
\end{equation}
Now, we proceed to compute the different terms in (\ref{kappasymmetricmatrix}). First, the antisymmetrized product of all gamma matrices gives
\begin{equation}
\gamma_{(7)} = M h^{\frac{1}{4}} e^{2f+2g} \sin  \alpha  \sin^2 \theta \Gamma_{012} \left( \frac{1}{x} \Gamma_3 +\theta'  \Gamma_8 \right) \Gamma_4 \left( \Gamma_6 - \cot \theta \Gamma_7 \right) \Gamma_9 \ d \zeta^7~~.
\label{antisymmetrizedgammamatrices}
\end{equation}
Using projections (\ref{kahlerproj}), (\ref{rotatedprojection0}) and (\ref{thirdprojection}),
\begin{equation}
\gamma_{(7)} \epsilon= M^{\frac{3}{2}} h^{\frac{1}{4}} e^{2f+2g} \sin  \alpha  \sin^2 \theta  \left( \frac{1}{x}  +\theta'  \cot \theta -\left( \theta'  - \frac{\cot \theta}{x}\right) \Gamma_{38} \right) \left( -1+ \frac{\Theta}{\sqrt{h}} \Gamma_{12} \Gamma_{11} \right) \epsilon \ d \zeta^7 ~~.
\label{antisymmetrizedgammamatrices1}
\end{equation}
This suggests to impose the following condition for the embedding function $\theta$,
\begin{equation}
\theta'  = \frac{\cot \theta}{x} ~~,
\label{susyembedding1}
\end{equation}
Then the expression (\ref{antisymmetrizedgammamatrices1}) simplifies to
\begin{equation}
\gamma_{(7)} \epsilon= M^{\frac{3}{2}} h^{\frac{1}{4}} e^{2f+2g} \sin  \alpha  \frac{1}{x}  \left( -1+ \frac{\Theta}{\sqrt{h}} \Gamma_{12} \Gamma_{11} \right) d \zeta^7 \ \epsilon ~~.
\label{antisymmetrizedgammamatrices2}
\end{equation}
The second term in (\ref{kappasymmetricmatrix}) gives
\begin{equation}
F \wedge \gamma_5 \Gamma_{11}=\big[ \gamma_{x_0x_1x_2}\left( F_{x\psi} \gamma_{\alpha \beta} - F_{x\beta} \gamma_{\alpha \psi}+F_{\alpha \beta} \gamma_{x \psi} \right)+F_{x_1 x_2} \gamma_{x_0 x \alpha \beta \psi } \big] \Gamma_{11}  d^7 \zeta ~~.
\label{Fgamma5term}
\end{equation}
The antisymmetric products of $\gamma$ matrices are:
\begin{align}
\gamma_{\alpha \beta}&= h^{\frac{1}{2}} e^{2f} \sin \alpha \sin \theta \left(  \Gamma_{46}-\frac{\cos \theta}{\sin \theta} \Gamma_{47} + \frac{\cos \alpha}{\sin \alpha} e^{g-f} \Gamma_{49}  \right)  \\
\gamma_{\alpha \psi}&= h^{\frac{1}{2}} e^{g+f} \sin \theta \ \Gamma_{49}  \\
\gamma_{x \psi}&= h^{\frac{1}{2}} e^{2g} \sin \theta  \left( \frac{1}{x} \Gamma_{39} + \theta' \Gamma_{89} \right)  ~.
\label{antisimetric_product_of_2_gamma_matrices}
\end{align}
Using projections (\ref{kahlerproj}) and (\ref{thirdprojection})
\bea
 && \gamma_{x_0x_1x_2}\left( F_{x\psi} \gamma_{\alpha \beta} - F_{x\beta} \gamma_{\alpha \psi}+F_{\alpha \beta} \gamma_{x \psi} \right) \Gamma_{11} \epsilon   \nonumber\\
&= & h^{\frac{1}{2}} \sin \alpha \left[ \sin \theta \left( a' e^{2f} + a e^{2g} \frac{1}{x} \right) \Gamma_{46} - \left( a' e^{2f} \cos \theta + a e^{2g} \sin \theta \theta' \right) \Gamma_{47} \right] \epsilon ~.
\label{Fgamma5term_simplification}
\eea
This expression vanishes if we impose (\ref{susyembedding1}) and a new condition on the function $a(x)$
\begin{equation}
a'  =- \frac{a}{x} e^{2g-2f} \ .
\label{susyembedding2}
\end{equation}
Let us now consider the last term in (\ref{Fgamma5term}). Using projections (\ref{kahlerproj}) and (\ref{thirdprojection})
\begin{equation}
 \gamma_{x_0 x \alpha \beta \psi }  \Gamma_{11}  \epsilon = h^{\frac{3}{4}} e^{2f+2g} \sin  \alpha  \sin^2 \theta  \left( \frac{1}{x}  +\theta'  \cot \theta - \left( \theta'  - \frac{\cot \theta}{x}\right) \Gamma_{38} \right) \Gamma_0 \Gamma_{11} \epsilon ~~,
\label{gamma122345b}
\end{equation}
and rewriting (\ref{rotatedprojection0}) as
\begin{equation}
\Gamma_0 \Gamma_{11} \epsilon = - \sqrt{M} \left( \frac{\Theta}{\sqrt{h}} +\Gamma_{12} \Gamma_{11} \right) \epsilon ~~,
\label{rotatedprojection3}
\end{equation}
and using (\ref{susyembedding1}) we obtain
\begin{equation} 
F \wedge \gamma_5 \Gamma_{11}=F_{x_1 x_2} \gamma_{x_0 x \alpha \beta \psi } \Gamma_{11} \epsilon \ d\zeta^7 =  - \Theta M^{\frac{3}{2}} h^{-\frac{1}{4}} e^{2f+2g} \sin  \alpha  \frac{1}{x}  \left( \frac{\Theta}{\sqrt{h}} +\Gamma_{12} \Gamma_{11} \right) \epsilon \ d\zeta^7~~.
\label{FwedgeFwedgegamma3epsilon}
\end{equation}
We move now to the next contribution in (\ref{kappasymmetricmatrix})
\begin{equation}
\frac{1}{2} F \wedge F \wedge \gamma_{(3)} \epsilon = d \zeta^7 \left[ F_{x_1 x_2}\gamma_{x_0} \left( F_{x \psi}\gamma_{\alpha \beta}-F_{x \beta} \gamma_{\alpha \psi}+ F_{\alpha \beta} \gamma_{x \psi} \right)+F_{x \psi}F_{\alpha \beta} \gamma_{x_0 x_1 x_2} \right]\epsilon~.
\label{third_term}
\end{equation}
The terms inside the round brackets cancel as in the previous step. Thus,
\begin{equation}
\frac{1}{2} F \wedge F \wedge \gamma_{(3)}\epsilon=a a' M h^{-\frac{3}{4}}\sin\alpha \Gamma_{012}\epsilon=a a' M^{\frac{3}{2}} h^{-\frac{3}{4}} \sin \alpha \left(-1+\frac{\Theta}{\sqrt{h}} \Gamma_{12}\Gamma_{11} \right) \epsilon \ d^7 \zeta~.
\label{third_term2}
\end{equation}
The last contribution is given by
\bea
 & & \frac{1}{6} F \wedge F \wedge F \wedge \gamma_{(1)} \Gamma_{11}\epsilon=\Theta M h^{-\frac{5}{4}} a a' \sin \alpha  \Gamma_{0}\Gamma_{11} \epsilon \ d\zeta^7 \nonumber\\
& = & -\Theta M^{\frac{3}{2}} h^{-\frac{5}{4}} a a' \sin \alpha  \left( \frac{\Theta}{\sqrt{h}}+\Gamma_{12}\Gamma_{11}\right) \epsilon \ d\zeta^7   \ .\label{FwedgeFwedgeFwedgegamma1epsilon}
\eea
Putting all together in (\ref{kappasymmetricmatrix})
\begin{equation}
\Gamma_{\kappa} \epsilon=- \frac{M^{\frac{1}{2}}h^{\frac{1}{4}}\sin \alpha }{\sqrt{-\det(\hat{g}+F)}} \left( e^{2f+2g}\frac{1}{x} + a a' h^{-1} \right) \epsilon~.
\label{kappasymmetricmatrix2}
\end{equation}
Finally, the root square of the determinant, after imposing the BPS embedding, reads
\begin{equation}
\sqrt{-\det(\hat{g}+F)}\big{|}_{BPS} = M^{\frac{1}{2}} h^{\frac{1}{4}} \sin \alpha \left(  e^{2f+2g} \frac{1}{x} +a a' h^{-1} \right)~,
\label{detgF}
\end{equation}
and we obtain that the embedding is kappa symmetric
\begin{equation}
\Gamma_{\kappa} \epsilon=-\epsilon~.
\label{kappasymmetricembedding}
\end{equation}
In summary, the kappa symmetry condition imposes equations (\ref{susyembedding1}) and (\ref{susyembedding2}) on the embedding functions $\theta(x)$ and $a(x)$, and we can integrate them to obtain:
\begin{equation}
   \theta (x) = \cos \left( \frac{x_*}{x} \right) ~~, ~~~~~~~~~~~~ a(x)=- Q e^{-\int_{x_*}^{\infty} \frac{e^{2f(x)-2g(x)}}{x}dx}~~,
\label{BPSembedding_equations_integrated}
\end{equation}
where $x_*$ is the location of the tip of the brane, and $-Q$ is the flux at $x_*$. The embedding equations are the same as for the commutative case consistent with the action of the TsT transformation, which gives the same equations but in a rotated manner.

\section{UV asymptotics of the background} \label{app:asymptotics}

UV asymptotics of metric functions are worked out in \cite{Bea:2013jxa}. The asymptotic forms we need are the following
\begin{align}
   h(x) &= \frac{L_{UV}^4}{\kappa^4 r_q^4} x^{-\frac{4}{b}} \left( 1+ \frac{h_2}{x^2} + \dots \right) \\
   e^{f(x)} &= \sqrt{q_0} \frac{\kappa r_q}{b} x^\frac{1}{b} \left( 1+ \frac{f_2}{x^2} + \dots \right) \\
   e^{g(x)} &= \frac{\kappa r_q}{b} x^\frac{1}{b} \left( 1+ \frac{g_2}{x^2} + \dots \right) \\
   e^{\phi(x)} &= e^{\phi_0} \left( 1+ \frac{\phi_2}{x_2} + \dots \right) \ ,
\end{align}
where the subleading coefficients given by the following functions of $b$
\begin{align}
   h_2 &= -\frac{2 (b-1) (3 b-5) (b (2 b (9 b-1)-55)+30)}{b^2 (2 b+3) ((b-13) b+15)} \\
   f_2 &= \frac{(b-1) (3 b-5) (b (12 b-25)+10)}{2 b^2 ((b-13) b+15)} \\
   g_2 &= \frac{(b-1) (3 b-5) (b (22 b-35)+10)}{2 b^2 ((b-13) b+15)} \\
   \phi_2 &= \frac{9 (b-1) (3 b-5) \left(3 b^2-5\right)}{b (2 b+3) ((b-13) b+15)} \ .
\end{align}


\begin{thebibliography}{99}

\bibitem{Snider}
H. S. Snyder, ``Quantized space-time,",  Phys.\ Rev. {\bf 71}, 38 (1947).

\bibitem{Seiberg:1999vs}
  N.~Seiberg and E.~Witten,
  ``String theory and noncommutative geometry,''
  JHEP {\bf 9909} (1999) 032
  [hep-th/9908142].
  

\bibitem{Douglas:2001ba}
  M.~R.~Douglas and N.~A.~Nekrasov,
  ``Noncommutative field theory,''
  Rev.\ Mod.\ Phys.\  {\bf 73} (2001) 977
  [hep-th/0106048].
  
\bibitem{Szabo:2001kg}
  R.~J.~Szabo,
  ``Quantum field theory on noncommutative spaces,''
  Phys.\ Rept.\  {\bf 378} (2003) 207
  [hep-th/0109162].


\bibitem{Ezawa}
Z. F. Ezawa, {\sl Quantum Hall Effects: Field Theoretical Approach and Related Topics }, World Scientific (2008). 


\bibitem{GirvinJach}
S. M. Girvin and T. Jach, 
``Formalism for the quantum Hall effect: Hilbert space for analytic functions,",
Phys. Rev. {\bf B 29} (1984) 5617.


\bibitem{Comtet:1999owa} 
 S. M. Girvin, ``The Quantum Hall Effect: novel excitations and broken symmetries"
 in  A.~Comtet, T.~Joliceur, S.~Ouvry and F.~David,
  ``Topological Aspects of Low-dimensional Systems: Proceedings, Les Houches Summer School of Theoretical Physics, Session 69 : Les Houches, France, July 7-31 1998,''

\bibitem{Fradkin:1991nr}
  E.~H.~Fradkin,
  ``Field Theories of Condensed Matter Physics,''
 Cambridge University Press (2013)


\bibitem{Zee:1996fe}
  A.~Zee,
  ``Quantum Hall fluids,''
  Lect.\ Notes Phys.\  {\bf 456} (1995) 99
  [cond-mat/9501022].


\bibitem{Tong:2016kpv}
  D.~Tong,
  ``Lectures on the Quantum Hall Effect,''
  arXiv:1606.06687 [hep-th].

\bibitem{Susskind:2001fb}
  L.~Susskind,
  ``The Quantum Hall fluid and noncommutative Chern-Simons theory,''
  hep-th/0101029.

\bibitem{Polychronakos:2001mi}
  A.~P.~Polychronakos,
  ``Quantum Hall states as matrix Chern-Simons theory,''
  JHEP {\bf 0104} (2001) 011
  [hep-th/0103013].

\bibitem{Barbon:2001dw}
  J.~L.~F.~Barbon and A.~Paredes,
  ``Noncommutative field theory and the dynamics of quantum Hall fluids,''
  Int.\ J.\ Mod.\ Phys.\ A {\bf 17} (2002) 3589
  [hep-th/0112185].
  
\bibitem{Fradkin:2002qw}
  E.~Fradkin, V.~Jejjala and R.~G.~Leigh,
  ``Noncommutative Chern-Simons for the quantum Hall system and duality,''
  Nucl.\ Phys.\ B {\bf 642} (2002) 483
  [cond-mat/0205653 [cond-mat.mes-hall]].
    
\bibitem{Cappelli:2006wa}
  A.~Cappelli and I.~D.~Rodriguez,
  ``Jain States in a Matrix Theory of the Quantum Hall Effect,''
  JHEP {\bf 0612} (2006) 056
  [hep-th/0610269].
  

\bibitem{Hashimoto:1999ut}
  A.~Hashimoto and N.~Itzhaki,
  ``Noncommutative Yang-Mills and the AdS / CFT correspondence,''
  Phys.\ Lett.\ B {\bf 465} (1999) 142
  [hep-th/9907166].
  


\bibitem{Maldacena:1999mh}
  J.~M.~Maldacena and J.~G.~Russo,
  ``Large N limit of noncommutative gauge theories,''
  JHEP {\bf 9909} (1999) 025
  [hep-th/9908134].
  



\bibitem{Li:1999am}
  M.~Li and Y.~S.~Wu,
  ``Holography and noncommutative Yang-Mills,''
  Phys.\ Rev.\ Lett.\  {\bf 84} (2000) 2084
  [hep-th/9909085].
 

\bibitem{Aharony:2008ug}
  O.~Aharony, O.~Bergman, D.~L.~Jafferis and J.~Maldacena,
  ``N=6 superconformal Chern-Simons-matter theories, M2-branes and their gravity duals,''
  JHEP {\bf 0810} (2008) 091
  [arXiv:0806.1218 [hep-th]].
  

\bibitem{Hohenegger:2009as}
  S.~Hohenegger and I.~Kirsch,
  ``A Note on the holography of Chern-Simons matter theories with flavour,''
  JHEP {\bf 0904} (2009) 129
  [arXiv:0903.1730 [hep-th]].
  
\bibitem{Gaiotto:2009tk}
  D.~Gaiotto and D.~L.~Jafferis,
  ``Notes on adding D6 branes wrapping RP**3 in AdS(4) x CP**3,''
  JHEP {\bf 1211} (2012) 015
  [arXiv:0903.2175 [hep-th]].

  
\bibitem{Veneziano:1976wm}
  G.~Veneziano,
  ``Some Aspects of a Unified Approach to Gauge, Dual and Gribov Theories,''
  Nucl.\ Phys.\ B {\bf 117} (1976) 519.
 

\bibitem{Nunez:2010sf}
  C.~Nunez, A.~Paredes and A.~V.~Ramallo,
  ``Unquenched Flavor in the Gauge/Gravity Correspondence,''
  Adv.\ High Energy Phys.\  {\bf 2010} (2010) 196714
  [arXiv:1002.1088 [hep-th]].
 


\bibitem{Conde:2011sw}
  E.~Conde and A.~V.~Ramallo,
 ``On the gravity dual of Chern-Simons-matter theories with unquenched
  flavor,''
  JHEP {\bf 1107}, 099 (2011)
  [arXiv:1105.6045 [hep-th]].


\bibitem{Jokela:2012dw}
  N.~Jokela, J.~Mas, A.~V.~Ramallo and D.~Zoakos,
  ``Thermodynamics of the brane in Chern-Simons matter theories with flavor,''
  JHEP {\bf 1302} (2013) 144
  [arXiv:1211.0630 [hep-th]].

\bibitem{Bea:2013jxa}
  Y.~Bea, E.~Conde, N.~Jokela and A.~V.~Ramallo,
  ``Unquenched massive flavors and flows in Chern-Simons matter theories,''
  JHEP {\bf 1312} (2013) 033
  [arXiv:1309.4453 [hep-th]].




\bibitem{Bea:2014yda}
  Y.~Bea, N.~Jokela, M.~Lippert, A.~V.~Ramallo and D.~Zoakos,
  ``Flux and Hall states in ABJM with dynamical flavors,''
  JHEP {\bf 1503} (2015) 009
  [arXiv:1411.3335 [hep-th]].
 



\bibitem{Martin:2017nhg}
  C.~P.~Martin, J.~Trampetic and J.~You,
  ``Quantum noncommutative ABJM theory: first steps,''
  arXiv:1711.09664 [hep-th].

\bibitem{imeroni}
  E.~Imeroni,
  ``On deformed gauge theories and their string/M-theory duals ,''
  JHEP 0810 (2008) 026
  [arXiv:0808.1271  [hep-th]].

\bibitem{Colgain:2016gdj}
  E.~\'O.~Colg\'ain and A.~Pittelli,
  ``A Requiem for $AdS_4 \times \mathbb{C} P^3$ Fermionic self-T-duality,''
  Phys.\ Rev.\ D {\bf 94} (2016) no.10,  106006
  [arXiv:1609.03254 [hep-th]].
  
\bibitem{Lunin:2005jy}
  O.~Lunin and J.~M.~Maldacena,
  ``Deforming field theories with U(1) x U(1) global symmetry and their gravity duals,''
  JHEP {\bf 0505} (2005) 033
  [hep-th/0502086].

  \bibitem{Alishahiha:1999ci}
  M.~Alishahiha, Y.~Oz and M.~M.~Sheikh-Jabbari,
  ``Supergravity and large N noncommutative field theories,''
  JHEP {\bf 9911} (1999) 007
  [hep-th/9909215].

\bibitem{Mateos:2002rx}
  T.~Mateos, J.~M.~Pons and P.~Talavera,
  ``Supergravity dual of noncommutative N=1 SYM,''
  Nucl.\ Phys.\ B {\bf 651} (2003) 291
  [hep-th/0209150].

  \bibitem{Bigatti:1999iz}
  D.~Bigatti and L.~Susskind,
  ``Magnetic fields, branes and noncommutative geometry,''
  Phys.\ Rev.\ D {\bf 62} (2000) 066004
  [hep-th/9908056].

\bibitem{Matusis:2000jf}
  A.~Matusis, L.~Susskind and N.~Toumbas,
  ``The IR / UV connection in the noncommutative gauge theories,''
  JHEP {\bf 0012} (2000) 002
  [hep-th/0002075].
  
\bibitem{Rey:1998ik}
  S.~J.~Rey and J.~T.~Yee,
  ``Macroscopic strings as heavy quarks in large N gauge theory and anti-de Sitter supergravity,''
  Eur.\ Phys.\ J.\ C {\bf 22} (2001) 379
  [hep-th/9803001].

\bibitem{Maldacena:1998im}
  J.~M.~Maldacena,
  ``Wilson loops in large N field theories,''
  Phys.\ Rev.\ Lett.\  {\bf 80} (1998) 4859
  [hep-th/9803002].

\bibitem{SheikhJabbari:1999vm}
  M.~M.~Sheikh-Jabbari,
 ``Open strings in a B field background as electric dipoles,''
  Phys.\ Lett.\ B {\bf 455} (1999) 129
  [hep-th/9901080].
  

\bibitem{Arean:2005ar}
  D.~Arean, A.~Paredes and A.~V.~Ramallo,
  ``Adding flavor to the gravity dual of non-commutative gauge theories,''
  JHEP {\bf 0508} (2005) 017
  [hep-th/0505181].

\bibitem{Haque:2009hz}
  S.~S.~Haque and A.~Hashimoto,
  ``Mass-spin relation for quark anti-quark bound states in non-commutative Yang-Mills theory,''
  Nucl.\ Phys.\ B {\bf 829} (2010) 555
  [arXiv:0903.4841 [hep-th]].
  
\bibitem{Landsteiner:2007bd}
  K.~Landsteiner and J.~Mas,
  ``The Shear viscosity of the non-commutative plasma,''
  JHEP {\bf 0707} (2007) 088
  [arXiv:0706.0411 [hep-th]].
  
\bibitem{Barbon:2008ut}
  J.~L.~F.~Barbon and C.~A.~Fuertes,
  ``Holographic entanglement entropy probes (non)locality,''
  JHEP {\bf 0804} (2008) 096
  [arXiv:0803.1928 [hep-th]].

\bibitem{Fischler:2013gsa}
  W.~Fischler, A.~Kundu and S.~Kundu,
  ``Holographic Entanglement in a Noncommutative Gauge Theory,''
  JHEP {\bf 1401} (2014) 137
  [arXiv:1307.2932 [hep-th]].

\bibitem{Karczmarek:2013xxa}
  J.~L.~Karczmarek and C.~Rabideau,
  ``Holographic entanglement entropy in nonlocal theories,''
  JHEP {\bf 1310} (2013) 078
  [arXiv:1307.3517 [hep-th]].

\bibitem{Ryu}
  S.~Ryu and T.~Takayanagi,
  ``Holographic derivation of entanglement entropy from AdS/CFT,''
  Phys.\ Rev.\ Lett.\  {\bf 96} (2006) 181602
  [arXiv:hep-th/0603001].
  S.~Ryu and T.~Takayanagi,
  ``Aspects of holographic entanglement entropy,''
  JHEP {\bf 0608} (2006) 045
  [arXiv:hep-th/0605073].

\bibitem{Jokela:2013qya}
  N.~Jokela, A.~V.~Ramallo and D.~Zoakos,
  ``Magnetic catalysis in flavored ABJM,''
  JHEP {\bf 1402} (2014) 021
  [arXiv:1311.6265 [hep-th]].


  

  


  
  
\bibitem{Witten:2003ya} 
  E.~Witten,
  ``SL(2,Z) action on three-dimensional conformal field theories with Abelian symmetry,''
  In *Shifman, M. (ed.) et al.: From fields to strings, vol. 2* 1173-1200
  [hep-th/0307041].
  
\bibitem{BurgessandDolan} 
  C.~P.~Burgess and B.~P.~Dolan,
  ``Particle vortex duality and the modular group: Applications to the quantum Hall effect and other 2-D systems,''
  Phys.\ Rev.\ B {\bf 63}, 155309 (2001)
  [hep-th/0010246];
  ``The Quantum Hall effect in graphene: Emergent modular symmetry and the semi-circle law,''
  Phys.\ Rev.\ B {\bf 76}, 113406 (2007)
  [cond-mat/0612269 [cond-mat.mes-hall]].

  

  \bibitem{Jokela:2013hta}
  N.~Jokela, G.~Lifschytz and M.~Lippert,
  ``Holographic anyonic superfluidity,''
  JHEP {\bf 1310} (2013) 014
  [arXiv:1307.6336 [hep-th]].

  
\bibitem{Brattan:2013wya}
  D.~K.~Brattan and G.~Lifschytz,
  ``Holographic plasma and anyonic fluids,''
  JHEP {\bf 1402} (2014) 090
  [arXiv:1310.2610 [hep-th]]. 

  \bibitem{Ihl:2016sop}
  M.~Ihl, N.~Jokela and T.~Zingg,
  ``Holographic anyonization: A systematic approach,''
  JHEP {\bf 1606} (2016) 076
  [arXiv:1603.09317 [hep-th]].
  
\bibitem{Bergman:2010gm}
  O.~Bergman, N.~Jokela, G.~Lifschytz and M.~Lippert,
  ``Quantum Hall Effect in a Holographic Model,''
  JHEP {\bf 1010} (2010) 063
  [arXiv:1003.4965 [hep-th]].


\bibitem{Jokela:2014wsa}
  N.~Jokela, G.~Lifschytz and M.~Lippert,
  ``Flowing holographic anyonic superfluid,''
  JHEP {\bf 1410} (2014) 21
  [arXiv:1407.3794 [hep-th]].

 
\bibitem{Jokela:2011eb}
  N.~Jokela, M.~J\"arvinen and M.~Lippert,
  ``A holographic quantum Hall model at integer filling,''
  JHEP {\bf 1105} (2011) 101
  [arXiv:1101.3329 [hep-th]].
  
\bibitem{Jokela:2011sw}
  N.~Jokela, M.~J\"arvinen and M.~Lippert,
  ``Fluctuations of a holographic quantum Hall fluid,''
  JHEP {\bf 1201} (2012) 072
  [arXiv:1107.3836 [hep-th]].
  
\bibitem{Jokela:2010nu}
  N.~Jokela, G.~Lifschytz and M.~Lippert,
  ``Magneto-roton excitation in a holographic quantum Hall fluid,''
  JHEP {\bf 1102} (2011) 104
  [arXiv:1012.1230 [hep-th]].

  
\bibitem{dilatinogravitino}
 L.~Martucci, J.~Rosseel, D.~Van~den~Bleeken and A.~Van~Proeyen
 ``Dirac actions for D-branes on backgrounds with fluxes,''
 Class. Quant. Grav. 22, 2745 (2005)
 [arXiv:0504041  [hep-th]].




\end{thebibliography}
\end{document}